%% file: main.tex
\documentclass[conference]{IEEEtran}

\input{packages}

\newcommand{\para}[1]{\noindent\textbf{{#1}}}

\def\ourmethod{SCAgent}
\def\SCAgentall{$\ourmethod{}_{\text{all}}$}
\def\SCAgentnew{$\ourmethod{}_{\text{new}}$}

\begin{document}
\IEEEoverridecommandlockouts

\title{Rethinking Side-Channel Analysis: Automated Discovery and Analysis of Side-Channel Leakage with LLM-Assisted Agents}

\author{
\IEEEauthorblockN{
Zhen Xu\IEEEauthorrefmark{1},
Zihao Wang\IEEEauthorrefmark{1},
Yuhua Sun\IEEEauthorrefmark{2},
XiaoFeng Wang\IEEEauthorrefmark{1}
}

\IEEEauthorblockA{
\IEEEauthorrefmark{1}Nanyang Technological University 
}

\IEEEauthorblockA{
\IEEEauthorrefmark{2}Xiangtan University
}

\thanks{The first two authors contributed equally to this work.}
\thanks{Correspondence to \href{mailto:zihao.wang@ntu.edu.sg}{zihao.wang@ntu.edu.sg}.}
}

\maketitle

\input{0_abstract.tex}
\IEEEpeerreviewmaketitle

\input{1_introduction}

\input{2_background}

\input{3_method}

\input{4_evaluation}

\input{5_discussion}

\input{6_conclusion}

\input{main.bbl}

\input{99_appendix}

\end{document}

%% file: packages.tex
\usepackage{tikz}
\usepackage{amsthm}
\usepackage{autobreak}
\allowdisplaybreaks
\usepackage{adjustbox}
\usepackage{booktabs}
\usepackage{makecell}
\usepackage{siunitx}
\usepackage{multirow}
\usepackage{balance}
\usepackage{longtable}
\usepackage{algorithm}
\usepackage{algpseudocode}

\algnewcommand\Input{\item[\textbf{Input:}]}
\algnewcommand\Output{\item[\textbf{Output:}]}

\usepackage{bm}

\usepackage{graphicx}
\usepackage{placeins}
\usepackage{amsmath}

\usepackage{booktabs}
\usepackage{tabularx}
\usepackage{listings}
\usepackage{fix-cm}
\usepackage{hyperref}
\usepackage{wasysym}

\usepackage{amssymb}
\usepackage{array}
\usepackage{tabularx}
\usepackage{caption}
\usepackage{xcolor}
\usepackage{bm}
\usepackage{amsthm}
\let\labelindent\relax
\usepackage{enumitem}

\newcommand{\pgftextcircled}[1]{
    \setbox0=\hbox{#1}%
    \dimen0\wd0%
    \divide\dimen0 by 2%
    \begin{tikzpicture}[baseline=(a.base)]%
        \useasboundingbox (-\the\dimen0,0pt) rectangle (\the\dimen0,1pt);
        \node[circle,draw,outer sep=0pt,inner sep=0.1ex] (a) {#1};
    \end{tikzpicture}
}

\theoremstyle{definition}
\theoremstyle{remark}

\newtheoremstyle{bfnote}%
  {}{}
  {\itshape}{}
  {\bfseries}{.}
  { }{\thmname{#1}\thmnumber{ #2}\thmnote{ (#3)}}
\theoremstyle{bfnote}

\hypersetup{
  colorlinks,
  linkcolor={blue!70!green},
  citecolor={green!70!blue},
  urlcolor={orange!70!red}
}

\usepackage{tcolorbox}
\tcbuselibrary{skins, breakable}

\newtcolorbox{prettybox}[2][]{
    enhanced,
    colframe=black!85,
    colbacktitle=black!85,
    coltitle=white,
    colback=white,
    fonttitle=\bfseries\large,
    arc=0.5mm,
    boxrule=0.5mm,
    titlerule=0pt,
    left=10pt, right=10pt, top=6pt, bottom=7pt,
    toptitle=5pt, bottomtitle=5pt,
    title={#2},
    width=0.85\linewidth,
    left skip=0.075\linewidth,
    breakable,
    parbox=false,
    #1
}

%% file: 0_abstract.tex
\begin{abstract}
Side-channel attacks exploit unintended information leakage from system behavior and continue to pose serious privacy risks in modern platforms. Despite extensive prior work, side-channel analysis remains largely manual and fragmented, typically assuming predefined target events and a fixed set of known channels. As systems and applications grow increasingly complex, several fundamental questions remain unanswered: which user or system events are sensitive in practice, how side channels associated with these events can be systematically discovered without exhaustive manual effort, and how their leakage can be analyzed at scale without prohibitive data collection and model training costs.

To address these questions, we present SCAgent, an automated framework for side-channel risk analysis. To identify sensitive targets beyond manually specified events, SCAgent performs agent-driven system exploration guided by LLM-based semantic reasoning. To systematically discover side channels while mitigating the risk of LLM hallucination, it reasons over system documentation and incorporates explicit verification to enforce semantic consistency, threat-model feasibility, and per-channel usability. To enable scalable analysis under limited data, SCAgent adopts a few-shot learning paradigm based on foundation models, avoiding the need to train bespoke models for each channel--event pair. To bridge the gap between raw time-series side-channel signals and tabular foundation models, SCAgent further introduces a time-shift--robust feature extraction layer that enables effective downstream analysis.

We instantiate SCAgent on iOS as a first step, focusing on OS-level side channels observable by unprivileged applications. Our evaluation spans standard benchmarks such as foreground app identification and website fingerprinting, as well as newly identified sensitive in-app activities in popular applications. Experimental results show that SCAgent is highly effective in uncovering previously unknown side-channel leakage and remains robust across different settings and evaluation conditions.
\end{abstract}

%% file: 1_introduction.tex
\section{Introduction}
Side-channel attacks exploit unintended information leakage from system behavior to infer sensitive user activities, posing serious privacy risks across diverse domains, including cryptographic systems, operating systems, browsers, and mobile platforms~\cite{Peeking, Memento, Screenmilker, Identity}. 
Recent studies further highlight the fundamental difficulty of mitigating these risks, because an attacker can aggregate evidence across multiple noisy channels, even when each reveals only a small amount of information, to infer sensitive activities with high confidence~\cite{ios, wang2023danger}. 
Despite decades of research, side-channel discovery and analysis remain largely ad hoc and manual. 
Existing work typically targets a small set of predefined events and channels identified through expert insight and system-specific investigation. 
This approach does not scale well to modern platforms, where software stacks expose numerous subtle attack surfaces and where sensitive user activities evolve rapidly with applications and usage patterns. 
As a result, it remains difficult to assess whether current defenses meaningfully reduce real-world side-channel risk, or whether undiscovered channels and unanticipated sensitive events continue to expose private information.

\para{Challenges in automated side-channel analysis.}
Automating side-channel discovery and analysis presents several fundamental challenges. 
First, identifying \emph{what} constitutes a sensitive event, that is, the target of an attack (e.g., user activities performed through an app), is non-trivial and highly context-dependent. 
Prior work almost exclusively assumes that target events are manually specified, relying on domain knowledge. 
This manual target-selection step limits scalability and may overlook application-specific privacy risks that are difficult to enumerate in advance.

Second, discovering \emph{which} channels may leak information about a given event is inherently hard. 
Traditional approaches typically rely on manual inspection or keyword-based searches over system documentation, which are coarse-grained and often inaccurate. 
Recent advances in large language models (LLMs)~\cite{chatgpt, perplexity, copilot, gemini-chat} offer new opportunities to reason over complicated system descriptions and generate candidate leakage hypotheses. 
However, directly applying LLMs to this task introduces a new challenge: LLM-generated hypotheses may hallucinate non-existent behaviors, rely on unrealistic assumptions, or propose channels that are infeasible under the target threat model~\cite{huang2025survey}.

Third, even when candidate channels are identified, systematically analyzing their leakage remains expensive. 
Collecting large datasets and training bespoke models for every channel--event pair quickly becomes impractical as the number of candidate channels and target events grows, posing a significant scalability barrier for realistic side-channel risk analysis.

\para{Our solution.}
In this paper, we present \ourmethod{}, a framework for automated side-channel risk discovery and analysis.
\ourmethod{} addresses the challenges above by decomposing the analysis into three stages.

First, \ourmethod{} identifies candidate sensitive events via agent-driven exploration guided by LLM-based semantic reasoning. 
Our design builds on existing mobile agents. 
A practical challenge, however, is that most mature mobile agents today target Android~\cite{deng2024mobile, wang2024mobileagentbench, rawles2025androidworld, chen2024spa, sun2025autoeval}. 
To validate the approach without engineering a new iOS-native agent from scratch, we focus on applications that have both Android and iOS versions. 
The Android-based agent is used only to discover semantically meaningful candidate events in an app's Android version. 
These events are then mapped to the corresponding iOS workflows, and all subsequent channel discovery, trace collection, and leakage evaluation are conducted on iOS. 
Thus, \ourmethod{} does not assume that Android system behavior transfers to iOS; it only leverages shared high-level user semantics in popular cross-platform apps.

Second, \ourmethod{} automates side-channel discovery by reasoning over system documentation to identify candidate channels. 
Although LLMs make such reasoning practical, naively relying on LLM-generated hypotheses risks hallucinations~\cite{huang2025survey}. 
Proposed channels may correspond to non-existent system behaviors, violate the assumed threat model, or prove uninformative for distinguishing target events. 
To mitigate these risks, \ourmethod{} treats LLMs as fallible hypothesis generators and constrains their outputs through explicit verification. 
The verifier checks whether each proposed channel is semantically consistent with documentation, feasible under the unprivileged-app threat model, and measurably useful for distinguishing target events. 
Verification outcomes are then fed back to refine subsequent proposals, enabling broad exploration while filtering spurious hypotheses.

Third, \ourmethod{} performs scalable leakage analysis under limited data. 
In realistic settings, collecting large volumes of labeled traces and training bespoke classifiers for every channel--event pair quickly becomes prohibitively expensive. 
\ourmethod{} addresses this challenge by adopting a few-shot learning paradigm built on foundation models. 
A key difficulty is that side-channel signals are inherently multivariate time series~\cite{multivariate}, often exhibiting temporal misalignment across executions. 
To bridge this gap, \ourmethod{} uses ROCKET~\cite{ROCKET} as a time-shift--robust feature extraction layer and applies a tabular foundation model, TabPFN~\cite{TabPFN_Nature}, for downstream classification. 
This combination enables data-efficient leakage analysis without training a bespoke model for each attack task.

\para{Our results.}
We instantiate \ourmethod{} on iOS as a concrete case study, focusing on OS-level side channels observable by unprivileged applications. 
Our main large-scale evaluation is conducted on an iPhone 13 running iOS 15.3 and an iPhone 14 running iOS 16.3. 
Using \ourmethod{}, we identify 39 previously unreported iOS OS-level measurement primitives spanning 24 system interfaces. 
We evaluate these channels on standard benchmarks, including foreground app identification and website fingerprinting, as well as on fine-grained sensitive in-app activities discovered by \ourmethod{}.

On standard benchmarks, \ourmethod{} achieves strong end-to-end attack performance. 
For foreground app identification, \ourmethod{} achieves 98.1\% accuracy over 100 apps, outperforming the strongest existing baseline by 4 percentage points. 
For website fingerprinting, \ourmethod{} achieves 96.7\% accuracy over 100 websites, improving over the strongest existing baseline by 3 percentage points. 
Importantly, these results are not solely driven by reusing channels known from prior work. 
When all channels reported by prior iOS side-channel studies are excluded from attack construction, \ourmethod{} still achieves 95.0\% accuracy on foreground app identification and 90.1\% accuracy on website fingerprinting. 
These results show that suppressing known channels alone may still leave newly discoverable channels exploitable.

Beyond standard benchmarks, \ourmethod{} enables systematic discovery and evaluation of fine-grained in-app activities that are difficult to enumerate manually. 
For example, \ourmethod{} identifies and evaluates sensitive activities related to travel planning, health tracking, and profile-related content creation. 
Such activities are application-specific, semantically meaningful, and often absent from prior side-channel benchmarks. 
This capability highlights the value of agent-assisted discovery: the main contribution is not merely improving fingerprinting accuracy on known tasks, but expanding side-channel analysis to previously unanticipated target events.

We further validate whether the observed leakage is specific to older iOS versions. 
In a focused experiment on iOS 26.4.2, \ourmethod{} achieves 96.5\% accuracy on website fingerprinting and 82.0\% accuracy on foreground app identification. 
The foreground-app validation uses a much smaller manually collected dataset, but the result still indicates that the leakage remains observable on a recent iOS version. 
Together, these findings demonstrate the effectiveness of \ourmethod{} and motivate more systematic auditing of OS-exposed interfaces.

\para{Responsible disclosure.}
In January 2026, we responsibly disclosed the newly identified side channels to Apple. 
As of the preparation of this manuscript, the Apple security team is actively examining the reported issues, and we remain in communication with them.

\para{Contributions.}
Our key contributions are as follows:

\noindent$\bullet$\textit{~New findings}.  
We demonstrate that agent-assisted side-channel discovery and exploitation is feasible on a modern mobile platform. 
\ourmethod{} uncovers previously unreported iOS OS-level measurement primitives and shows that they remain exploitable even after channels known from prior work are excluded. 
Our results suggest that side-channel risk is not limited to a fixed set of manually known APIs, but can evolve as automated tools explore larger interface spaces.

\noindent$\bullet$\textit{~New techniques}.  
We propose \ourmethod{}, a structured framework for side-channel risk analysis that combines agent-driven sensitive-event discovery, LLM-assisted channel proposal with explicit verification, and foundation-model-based leakage analysis. 
The key design choice is to treat LLMs as fallible hypothesis generators and to constrain their outputs through documentation consistency checks, threat-model feasibility checks, and lightweight distinguishability tests.

\noindent$\bullet$\textit{~Empirical evaluation}.  
We evaluate \ourmethod{} on real iOS devices across foreground app identification, website fingerprinting, and fine-grained in-app activity inference. 
The results show that newly discovered channels can support high-accuracy inference, and a focused validation on iOS 26.4.2 further indicates that such leakage remains observable on a recent iOS version.

%% file: 2_background.tex
\section{Background and Related Work}
\label{sec:back}

\subsection{Large Language Models}
Large language models (LLMs)~\cite{delgado2025gpt, liu2025exploring, yuan2025incident, yuanyuan2025annotation, kim2025aiorchestration} have become a general-purpose reasoning substrate, showing strong capabilities in understanding, synthesizing, and generating natural language across diverse tasks. 
Recent multimodal LLMs further extend this capability to heterogeneous inputs such as images, structured data, and system-level descriptions, lowering the barrier for applying high-level semantic reasoning to complex, multi-step problems.
From a security perspective, LLMs are attractive for automating tasks that traditionally require substantial human expertise. 
Recent studies have explored their use in penetration testing~\cite{deng2023pentestgpt}, protocol fuzzing~\cite{meng2024large}, CAPTCHA solving~\cite{deng2025oedipus}, and blockchain analysis~\cite{he2024large}. 
By translating high-level objectives into concrete action sequences, LLMs offer a new paradigm for scalable and adaptive security analysis. 
These capabilities suggest their potential for side-channel discovery and analysis, although this direction has not been systematically studied.

\subsection{Side-Channel Attacks}
Unlike attacks that exploit vulnerabilities to directly access sensitive data, side-channel attacks infer such information from unintended side effects produced during system or application execution. 
These side effects may appear benign, but can reveal secret data or user activities. 
Common sources include hardware components (e.g., caches), sensors, and system APIs exposed to third-party applications for querying device or OS status, commonly referred to as OS-level side channels. 
On mobile platforms such as iOS and Android, leakage of user activity information is particularly sensitive, as it can enable phishing, behavioral profiling, and other privacy attacks~\cite{Pardon}.

\para{OS-level side-channel attacks.}
We focus on OS-level side-channel attacks on iOS as a first step. 
Prior studies show that operating systems often inadequately control information exposure through such channels. 
A well-studied example is \texttt{procfs} on UNIX-like systems, including Android, which exposes per-process kernel statistics such as CPU, memory, and network usage, and has enabled a wide range of side-channel attacks~\cite{Identity, Peeping, Screenmilker, Memento, Pardon, Peeking}. 
In contrast, iOS does not provide \texttt{procfs} and exposes only aggregated, system-wide statistics to third-party apps. 
Although intended to be safe, these statistics remain highly informative when analyzed collectively, allowing user activities to be inferred from memory, network, and file-system signals without permissions~\cite{ios, wang2023danger}. 
Apple has responded with mitigations such as reducing update frequency and lowering the precision of certain statistics~\cite{CVE-2017-13852, CVE-2017-13873}.

\subsection{Automated and Systematic Side-Channel Discovery}
Prior work has explored automated or semi-automated side-channel discovery in specific domains. 
ProcHarvester~\cite{spreitzer2018procharvester} automatically analyzes \texttt{procfs} leakage on Android, focusing on kernel-exposed process information. 
AndroTIME~\cite{palfinger2020androtime} identifies timing side channels in Android APIs and demonstrates their use for application inference. 
Cache template attacks~\cite{gruss2015cache} automate the discovery of action-specific cache activity at the microarchitectural level. 
These works demonstrate the value of systematic discovery, but differ from \ourmethod{} in scope and abstraction. 
\ourmethod{} targets iOS OS-level interfaces under an unprivileged-app threat model, uses agent-driven semantic exploration to discover sensitive target events, and combines LLM-based channel proposal with explicit feasibility and distinguishability verification. 
Thus, \ourmethod{} complements prior automated discovery efforts by moving toward end-to-end agent-assisted discovery of both sensitive events and exploitable OS-level channels.

\subsection{Threat Model}
\label{subsec:threat}

We consider an adversary whose goal is to infer sensitive events on a target iOS device, such as app launches, website visits, and in-app activities. 
The adversary does so by analyzing information leaked through API returns exposed to applications by the operating system.
The adversary deploys an unprivileged attack app on the target device. 
The app masquerades as a benign application with a legitimate reason to run in the background (e.g., an audio player), but requests no permissions requiring explicit user consent. 
Beyond standard app installation, the adversary assumes no additional privileges, jailbreak, or system compromise.
From the defender's perspective, the goal is to limit information leakage from OS-exposed interfaces queryable by third-party applications. 
The defender operates at the OS level and may apply standard mitigations to API outputs, including disabling known channels, reducing update frequency, reducing precision, or adding noise.

\para{Operational scope of OS-level side channels.}
In this work, ``OS-level side channel'' is an operational definition tied to the attack surface explored by \ourmethod{}. 
We use this term for any leakage primitive whose observable signal is obtained through OS-exposed interfaces available to an unprivileged application under the standard app sandbox. 
The focus is therefore on what the attacker can measure at the software interface level, not on the lowest-level physical cause of the signal.

Under this definition, a channel may still be influenced by lower-level system behavior, such as cache contention, GPU scheduling, filesystem activity, or other shared-resource effects. 
Such influence does not move the channel outside our scope as long as the attacker does not directly access microarchitectural state, perform cache-set profiling, or rely on hardware-specific primitives such as Flush+Reload~\cite{yarom2014flush} or Prime+Probe~\cite{liu2015last}. 
Instead, the attacker only observes values or latencies returned by OS-accessible interfaces. 
Thus, the boundary of this work is defined by the observable attack primitive and the adversary's privileges, rather than by whether the signal is ultimately affected by lower-level resource contention.

Accordingly, we do not study attacks that require direct access to microarchitectural effects~\cite{Occupancy, Sandbox, fish}, power consumption~\cite{Chen2017POWERFULMA, Michalevsky2015PowerSpyLT}, electromagnetic emissions~\cite{Offline, ECDSA}, or other physical side channels. 
Such attacks rely on different measurement primitives and adversary capabilities, and would require different verifier rules and data-collection procedures.

\para{Scope of generality.}
\ourmethod{} is general at the workflow level: it separates target discovery, channel discovery, and leakage analysis. 
However, the concrete instantiation studied here is designed for the OS-exposed attack surface defined above. 
Extending the workflow to physical or microarchitectural side channels would require new measurement primitives, noise models, and feasibility checks tailored to those settings.

%% file: 3_method.tex
\section{Design of \ourmethod{}}
\label{sec:method}

\subsection{Overview}

We propose \ourmethod{}, a framework for automated side-channel risk analysis. 
At a high level, \ourmethod{} decomposes the analysis into three stages: 
(i) automated discovery of sensitive events, 
(ii) automated discovery of side channels, and 
(iii) automated side-channel analysis. 
The workflow is general: it separates the question of \emph{what} events may be privacy-sensitive, \emph{which} observable channels may leak information about them, and \emph{how} the leakage can be analyzed under limited data. 
In this paper, we instantiate this workflow for iOS OS-level side channels observable by unprivileged applications.

\para{Stage 1: Automated discovery of sensitive events.}
The first stage identifies candidate target events. 
Prior side-channel studies usually assume that the target events are manually specified, relying on expert knowledge and limiting scalability. 
This assumption is restrictive because sensitive events are often application-specific, context-dependent, and may evolve with app functionality and user behavior.

\ourmethod{} addresses this challenge through agent-driven exploration. 
A mobile agent interacts with an application as a benign user would, executes user-visible actions, and records the resulting application behaviors. 
An LLM then reasons about the semantic sensitivity of these behaviors, for example whether they reveal user intent, personal attributes, or private states. 
The output of this stage is a set of candidate sensitive events. 
These events are not assumed to be exploitable; they are only prioritized for later side-channel discovery and leakage analysis.
In our iOS instantiation, a practical challenge is that mature mobile agents are predominantly available for Android, while comparable iOS agents remain limited. 
We therefore focus on applications that have both Android and iOS versions. 
The Android-based agent is used to discover semantically meaningful candidate events in the Android version of an app, and the corresponding iOS workflows are then validated and analyzed in subsequent stages. 
This design uses Android only for semantic event discovery. 
All channel discovery, trace collection, and leakage evaluation are conducted on iOS.

\para{Stage 2: Automated discovery of side channels.}
Given a target event, the second stage discovers candidate side channels that may leak information about it. 
Manual inspection and keyword-based searches over system interfaces are difficult to scale and may miss non-obvious leakage sources. 
\ourmethod{} instead uses an iterative proposer--verifier process. 
The proposer reasons over system documentation, architectural descriptions, and an initial candidate channel database to generate hypotheses about observable signals correlated with the target event. 
Because LLM-generated hypotheses may be hallucinated, infeasible, or uninformative, the verifier checks each proposed channel before it is added to the database.

The verifier performs three checks. 
First, it checks semantic consistency with system documentation, ensuring that the proposed channel corresponds to documented or observable system behavior. 
Second, it checks feasibility under the threat model, ensuring that the channel can be measured by an unprivileged app under the standard sandbox and background-execution constraints. 
Third, it performs a lightweight usability check to test whether the channel exhibits measurable distinguishability for the target event. 
Rejected proposals are summarized and fed back to the proposer, enabling later iterations to avoid similar invalid hypotheses.

Verified channels are stored in a persistent candidate channel database. 
This database is a working set of plausible channels, not the final attack channel set. 
It may be initialized from any seed source, such as known side-channel literature, manually provided candidate interfaces, or an empty set. 
In our iOS implementation, we initialize the database using channels reported by prior iOS side-channel studies, which provides a realistic prior knowledge base and allows us to explicitly separate rediscovered channels from newly proposed ones.

\para{Stage 3: Automated side-channel analysis.}
The final stage analyzes whether the discovered channels support practical inference. 
A key challenge is scalability. 
Collecting large labeled datasets and training bespoke classifiers for every channel--event pair quickly becomes expensive as the number of candidate channels and target events grows. 
For example, if an attacker considers 1{,}000 target classes and collects 1{,}000 samples per class, this already requires one million labeled executions for a single channel. 
Across tens or hundreds of channels, the cost becomes unrealistic for systematic analysis.

\ourmethod{} addresses this challenge with a few-shot analysis pipeline based on foundation models. 
Side-channel traces are multivariate time series and often exhibit temporal misalignment across executions, while current foundation models for time-series analysis remain limited in this setting. 
\ourmethod{} therefore converts raw traces into a tabular representation: ROCKET extracts time-shift--robust features from multivariate traces, PCA reduces feature dimensionality, and TabPFN performs few-shot classification without task-specific model training. 
This design reduces labeled-data requirements and avoids training a bespoke model for each channel--event pair.

\para{Prompt structure and automation.}
\ourmethod{} should be understood as an agent-assisted automated framework rather than an unconstrained open-ended discovery system. 
Its prompts encode high-level task structure, including the threat model, feasibility constraints, and the need to reason about OS-exposed resources. 
They do not enumerate the final discovered channels or specify target event--channel pairs. 
This structure is necessary because unconstrained LLM outputs often include hallucinated APIs, privileged interfaces, or channels unavailable under background execution. 
Thus, prompt design specifies the analysis constraints, while concrete hypotheses, rejection outcomes, selected channels, and final attacks are generated by the pipeline.

\subsection{Automated Discovery of Sensitive Events}
\label{subsec:Discovery of Sensitive Events}

A sensitive target event is a user-visible behavior that may reveal user intent, personal attributes, or private states. 
Existing side-channel studies typically assume that such events are manually specified. 
This limits scalability and introduces analyst bias, because privacy-sensitive behaviors may be application-specific and difficult to enumerate. 
\ourmethod{} instead discovers candidate sensitive events through agent-driven exploration.

\para{Mobile-agent exploration.}
The mobile agent explores an application by executing user-level actions such as clicks, tab switches, form-entry actions, and navigation actions. 
The action space is generated from scratch by recursively parsing the UI state exposed to the agent; it is not initialized from a manually curated list of sensitive actions. 
During exploration, the agent records the interaction context, including the application category, the executed action sequence, and the resulting UI behavior.

\para{Sensitive event identification.}
For each observed behavior, \ourmethod{} asks an LLM to assess whether the event is potentially sensitive in the context of the application and the executed action sequence. 
The LLM is given a structured description of the interaction, including the application category, the action sequence, and a natural-language description of the resulting behavior. 
Events judged to reveal user intent, personal attributes, or private states are added to the candidate sensitive-event set $\mathcal{E}_{\mathrm{sensitive}}$. 
Importantly, this stage does not assume that every event in $\mathcal{E}_{\mathrm{sensitive}}$ is exploitable through side channels. 
It only identifies privacy-relevant events that warrant further investigation. 
By separating sensitive-event discovery from leakage analysis, \ourmethod{} can systematically explore potential privacy risks before testing whether they are inferable.

\para{Scope of Android-to-iOS transfer.}
Because existing mobile agents are predominantly developed for Android, our implementation explores Android versions of cross-platform applications and then maps the discovered semantic events to corresponding iOS workflows. 
This use of Android is limited to semantic event discovery rather than side-channel measurement or attack validation. 
Android is used only to discover candidate user-level activities, such as selecting a pregnancy-related option or entering a travel-related workflow. 
All channel discovery, trace collection, and leakage evaluation are conducted on iOS. 
Thus, \ourmethod{} does not assume that Android system behavior transfers to iOS; it only assumes that many popular cross-platform apps preserve high-level user semantics across platforms.

\subsection{Automated Discovery of Side Channels}

Given a target event $E$, \ourmethod{} discovers candidate side channels through an iterative proposer--verifier process. 
Let $\mathcal{D}$ denote the system information available to the proposer, such as documentation, architectural descriptions, and developer-facing specifications. 
Let $\mathcal{C}_{\mathrm{seed}}$ denote an optional seed set of candidate channels. 
This seed set is part of the framework interface: it can be empty, manually provided, or initialized from prior knowledge. 
The candidate channel database is initialized as
\begin{equation}
\mathcal{C}_{\mathrm{db}}^{(0)} = \mathcal{C}_{\mathrm{seed}} .
\label{eq:db-init}
\end{equation}

\para{Seed channel database.}
The seed database serves two purposes. 
First, it provides examples that help the proposer reason about plausible leakage primitives in the target system. 
Second, it provides a reference point for distinguishing rediscovered channels from newly proposed ones. 
The seed set is not required by the framework: \ourmethod{} can in principle start from an empty database. 
In our iOS instantiation, we set $\mathcal{C}_{\mathrm{seed}}$ to the OS-level channels reported by prior iOS side-channel studies~\cite{ios, wang2023danger}. 
This is an implementation choice that provides a realistic prior knowledge base and makes rediscovery explicit. 
To isolate the contribution of newly discovered channels, we also evaluate a stricter setting in which all channels reported by prior work are excluded (see \autoref{subsec:main}).

\para{Proposer.}
At iteration $t$, the proposer $M_p$ generates a batch of $B$ candidate channels conditioned on the target event $E$, the system information $\mathcal{D}$, and the current database $\mathcal{C}_{\mathrm{db}}^{(t)}$:
\begin{equation}
\mathcal{C}_{\mathrm{prop}}^{(t)}
=
\mathcal{P}_{M_p}(E,\mathcal{D},\mathcal{C}_{\mathrm{db}}^{(t)}).
\label{eq:proposal}
\end{equation}
Each proposed channel describes an observable signal and a measurement mechanism. 
The batch-based design allows \ourmethod{} to explore multiple hypotheses per iteration, while later verifier feedback guides subsequent proposals.

\para{Verifier.}
The verifier $M_v$ filters proposed channels before they are added to the database. 
Given a proposal batch, it returns the subset of valid channels:
\begin{equation}
\mathcal{C}_{\mathrm{valid}}^{(t)}
=
\mathcal{V}_{M_v}(\mathcal{C}_{\mathrm{prop}}^{(t)},E,\mathcal{D}).
\label{eq:verification}
\end{equation}
For each candidate, the verifier checks documentation consistency, threat-model feasibility, and lightweight distinguishability. 
The first two checks ensure that the channel corresponds to plausible OS behavior and can be measured by the adversary. 
The third check filters channels that are feasible but uninformative for the target event.

For distinguishability, let $\mathrm{Acc}(c,E)$ denote the accuracy obtained from a lightweight evaluation of channel $c$ on event $E$, and let $\mathrm{Acc}_{\mathrm{rand}}(E)$ denote random-guessing accuracy. 
A channel is accepted only if
\begin{equation}
\mathrm{Acc}(c,E) \geq \tau_{\mathrm{dist}} \cdot \mathrm{Acc}_{\mathrm{rand}}(E),
\label{eq:distinguishability}
\end{equation}
where $\tau_{\mathrm{dist}}>1$ is a predefined threshold, set to $2$ by default.

Accepted channels are added to the database:
\begin{equation}
\mathcal{C}_{\mathrm{db}}^{(t+1)}
=
\mathcal{C}_{\mathrm{db}}^{(t)}
\cup
\mathcal{C}_{\mathrm{valid}}^{(t)} .
\label{eq:db-update}
\end{equation}
Rejected proposals are discarded, but their failure patterns are summarized as feedback to the proposer. 
This feedback includes, for example, rejected channel patterns, violated threat-model assumptions, and characteristics of validated channels. 
The process continues until a predefined budget is reached or the acceptance rate of a proposal batch falls below $\tau_{\mathrm{stop}}$, set to $0.2$ by default.

\para{Lightweight usability check.}
The usability check is a pre-screening step, not the final attack evaluation. 
For each proposed channel, \ourmethod{} collects a small calibration set from the training split only and never uses held-out testing traces during verification. 
It records short traces for the target event classes and computes $\mathrm{Acc}(c,E)$ using this training-only calibration data. 
This prevents clearly uninformative or infeasible proposals from entering the channel database while avoiding leakage from the final held-out evaluation.

\noindent\autoref{alg:stage2} (Appendix~\ref{subsec:alg}) presents the procedure for Stage~2.

\subsection{Automated Side-Channel Analysis}

After side channels are discovered, \ourmethod{} analyzes their leakage using a data-efficient foundation-model pipeline. 
The goal is to construct an attack classifier for a target event without collecting large datasets or training a bespoke model for every channel--event pair.

\para{Tabular foundation model for side-channel analysis.}
Side-channel traces are usually time-series signals, and existing time-series foundation models remain limited for short, noisy, and temporally misaligned traces. 
We therefore transform traces into tabular features and use TabPFN~\cite{TabPFN_Nature}, a transformer-based tabular foundation model pre-trained on large synthetic datasets~\cite{hollmann2022tabpfn, TabPFN_Nature}. 
Given a labeled context set and an unlabeled test instance, TabPFN performs in-context prediction without task-specific gradient updates. 
This makes it suitable for few-shot side-channel analysis, where labeled traces are expensive to collect.

\para{Channel selection.}
Before feature extraction, \ourmethod{} selects a compact subset of verified channels for the target event. 
Monitoring all verified channels may be inefficient and less stealthy, because it increases collection overhead and the observable footprint of the attack. 
Given the candidate database $\mathcal{C}_{\mathrm{db}}$ and target event $E$, an LLM-based selector $M_s$ outputs
\begin{equation}
\mathcal{C}_{\mathrm{sel}}
=
\mathcal{S}_{M_s}(E,\mathcal{C}_{\mathrm{db}}),
\qquad
\mathcal{C}_{\mathrm{sel}} \subseteq \mathcal{C}_{\mathrm{db}} .
\label{eq:channel-selection}
\end{equation}
The selector prioritizes channels that are likely to distinguish the classes induced by $E$ while avoiding redundant signals. 
The selected subset is then passed to feature extraction.

Importantly, initializing $\mathcal{C}_{\mathrm{db}}$ with $\mathcal{C}_{\mathrm{seed}}$ does not give seed channels privileged status in attack construction. 
Seed channels and newly proposed channels are placed in the same candidate pool and are selected only if $M_s$ judges them useful for the specific target event. 
Thus, seed channels are not automatically included in $\mathcal{C}_{\mathrm{sel}}$, and \ourmethod{} does not simply reuse all channels from prior work. 
For fairness, our evaluation also considers a stricter setting in which all channels reported by prior iOS studies are removed before attack construction.

\para{Feature extraction and selection.}
Let $\mathcal{C}_{\mathrm{sel}}$ be the selected channel set. 
A collected trace is represented as a multivariate time series
\begin{equation}
X=[X_1,X_2,\ldots,X_T],
\qquad
X_t \in \mathbb{R}^{|\mathcal{C}_{\mathrm{sel}}|},
\end{equation}
where each dimension corresponds to one selected side channel. 
Given a dataset
\begin{equation}
\mathcal{D}_{\mathrm{trace}}
=
\{(X^1,Y^1),(X^2,Y^2),\ldots,(X^N,Y^N)\},
\end{equation}
the goal is to infer the event label $Y$ from the trace $X$.

Raw traces are not suitable for TabPFN because identical events may be temporally misaligned across executions. 
\ourmethod{} therefore applies ROCKET to transform each trace into a fixed-length, time-shift--robust feature vector. 
ROCKET applies $L$ randomly generated convolutional kernels and extracts simple summary statistics, such as maximum response and the proportion of positive values, yielding
\begin{equation}
\phi(X) \in \mathbb{R}^{2L}.
\end{equation}
Because $2L$ can be large, \ourmethod{} applies PCA to the ROCKET feature matrix and obtains a compact representation
\begin{equation}
\tilde{\phi}(X)=\psi(\phi(X))\in\mathbb{R}^{d},
\qquad
d \ll 2L .
\label{eq:reduced-feature}
\end{equation}

\para{Inference with TabPFN.}
Applying the feature pipeline to all traces yields a tabular dataset
\begin{equation}
\tilde{\mathcal{D}}
=
\{(\tilde{\phi}(X^1),Y^1),(\tilde{\phi}(X^2),Y^2),\ldots,(\tilde{\phi}(X^N),Y^N)\}.
\end{equation}
Given a labeled context set $\tilde{\mathcal{D}}^{\mathrm{train}}\subset\tilde{\mathcal{D}}$ and a test trace $X^\ast$, TabPFN outputs
\begin{equation}
P(Y^\ast \mid \tilde{\phi}(X^\ast),\tilde{\mathcal{D}}^{\mathrm{train}})
\end{equation}
in a single forward pass. 
This inference paradigm reduces labeled-data requirements, accommodates feature-distribution variation across channels or devices, and avoids training separate models for many channel--event pairs.
The output is a classifier over event labels and a quantitative measure of event distinguishability. 
Together, channel selection, ROCKET-based feature extraction, PCA, and TabPFN enable efficient and scalable side-channel analysis under limited data.

\noindent\autoref{alg:stage3} (Appendix~\ref{subsec:alg}) presents the procedure for Stage~3.

%% file: 4_evaluation.tex
\section{Evaluation and Findings}
\label{sec:eval}

\subsection{Experimental Setup}
\label{subsec:setup}

\para{Devices and evaluation tasks.}
All experiments were conducted on an iPhone~13 running iOS~15.3 and an iPhone~14 running iOS~16.3.
Our main large-scale measurements use an iPhone 14 running iOS 16.3 because newer devices, such as iPhone 15 and later models, require iOS 17 or above, for which stable jailbreak support is not available for our large-scale measurement workflow. 
Although iPhone 14 can be upgraded to newer iOS versions, doing so would remove the jailbreak capability needed for controlled trace collection. 
Importantly, jailbreak is used only to facilitate data collection and experimental instrumentation. 
The attack itself relies solely on OS-exposed interfaces available to ordinary applications and does not require jailbreak, elevated privileges, or system compromise. 
Using older but jailbreakable setups for measurement is common in this line of side-channel research~\cite{ios, wang2023danger}.

For evaluation, we consider two classic benchmarks that are widely used in prior work~\cite{ios, wang2023danger}: \emph{foreground app identification} and \emph{website fingerprinting}. 

For foreground app identification, we use the top 100 most-reviewed free apps from the iOS App Store as target applications. 
The number of reviews provides a stable proxy for broad deployment and long-term user exposure, whereas instantaneous ranking lists can fluctuate significantly over time and across regions. 
We focus on free apps because they are easily accessible to both benign users and potential attackers, and because they cover a broad range of common app categories. 
For cross-platform sensitive-event discovery, we only retain apps that have both Android and iOS versions, so that the Android-based exploration stage can identify candidate semantic events later validated on iOS. 
The attacker's goal is to determine whether any of these target apps is launched within a given time window. 

For website fingerprinting, we use the top 100 websites ranked by Moz based on their overall popularity\footnote{\url{https://moz.com/top500}}. 
All evaluations are conducted using the default browser, Safari, and the attacker aims to infer whether any target website is visited within a given time window.

In addition to these benchmarks, we further identify new targets within the same set of top 100 most-reviewed free apps. 
The attacker aims to infer the occurrence of such sensitive in-app activities and monitor them within a given time window.

\para{Pipeline statistics.}
Across the evaluated applications, Stage~1 analyzed 1,433 candidate user actions generated through recursive UI parsing and semantic exploration. 
From these actions, \ourmethod{} identified candidate sensitive events, from which we selected 18 representative events for detailed side-channel evaluation. 
In Stage~2, the proposer generated candidate channels in batches of size $B=10$, and the verifier accepted 39 channels under the default Gemini 3 Pro configuration. 
These 39 channels span 24 OS-level interfaces. 
The rejected proposals were categorized into background-execution violations, sandbox violations, lack of distinguishability, and contradictory documentation evidence, as analyzed in Section~\ref{sec:llm_choice}.

\para{\ourmethod{} configuration.}
As described in \autoref{sec:method}, \ourmethod{} consists of three stages: 
(i) automated discovery of sensitive events, 
(ii) automated discovery of side channels, and 
(iii) automated side-channel analysis.
For the automated discovery of sensitive events, we use Mobile-Agent-E~\cite{wang2025mobile} with Gemini~3~Pro as the backend LLM.
As described in \autoref{subsec:Discovery of Sensitive Events}, this stage focuses on applications that have both iOS and Android versions to enable cross-platform exploration.
All exploration is conducted on the publicly available Android Emulator\footnote{\url{https://developer.android.com/studio}}.
The prompts used in this stage, as well as those used in subsequent stages, are provided in \autoref{sec:Prompts}.

For the automated discovery of side channels, the prior channel set $\mathcal{C}_{\text{prior}}$ is initialized using channels identified in prior work~\cite{ios, wang2023danger}.
Both the proposer model $M_p$ and the verifier model $M_v$ are instantiated using Gemini~3~Pro.
The batch size is set to $B=10$.
By default, the distinguishability threshold is set to $\tau_{\text{dist}}=2$, and the stopping threshold is set to $\tau_{\text{stop}}=0.2$.

For automated side-channel analysis, the channel selection model $M_s$ is set to Gemini~3~Pro by default.
The querying frequency for the selected channels is set to $T=100$.
Each target event is triggered 50 times, resulting in $N=50$ samples per event.
These samples are split into training and testing sets using a split ratio of $\alpha=0.8$.
For feature extraction, we use ROCKET with $L=10{,}000$ random kernels.
We then apply PCA for feature selection, reducing the feature dimensionality to 250.
Finally, for tabular classification, we use TabPFN with its default configuration~\cite{TabPFN_Nature}.

\para{App Store vetting.}
As part of our study, we submitted a monitoring application implementing the attacks enabled by \ourmethod{} to the App Store to assess whether such behavior would be flagged during review.
The application was presented as an audio player and requested the \emph{Audio Background Mode} permission to enable background execution.
It successfully passed the App Store review process, suggesting that software capable of launching these attacks may not be detected under current vetting practices.

We took several safeguards to avoid user harm. 
The application did not collect, transmit, or store any user data. 
Although it passed the App Store review process, we did not proceed with releasing or listing it on the App Store, which requires a separate developer action after approval. 
As a result, the application was never publicly available and received no downloads, so no users were exposed to the monitoring functionality. 
We also responsibly disclosed the discovered channels to Apple before public release and did not release one-click deployment artifacts for unsupervised use against real users.

\para{Additional recent-iOS validation.}
To examine whether the observed leakage is specific to older iOS versions, we additionally conducted a focused validation on iOS 26.4.2. 
For website fingerprinting, we collected 4,800 training traces and 1,200 testing traces, corresponding to 48 and 12 traces per website over 100 websites, respectively. 
For foreground app identification, due to the lack of automated trace collection on the recent iOS setup, we manually collected 400 training traces and 100 testing traces, corresponding to 4 and 1 trace per app over 100 apps, respectively. 
Both experiments follow the same attack settings as their corresponding main benchmarks, except that they are conducted on the recent iOS version.

\para{Evaluation metric.}
Unless otherwise stated, ``accuracy'' refers to end-to-end attack accuracy on the held-out testing set after \ourmethod{} has completed event discovery, channel discovery, channel selection, feature extraction, and classification. 
For a $K$-class target event with testing set $\mathcal{D}_{\mathrm{test}}=\{(X_i,y_i)\}_{i=1}^{n}$, we compute
\[
\mathrm{Acc}
=
\frac{1}{n}
\sum_{i=1}^{n}
\mathbf{1}\!\left[f_{SCAgent}(X_i)=y_i\right],
\]
where $f_{SCAgent}$ is the final classifier constructed from the event--channel pairs produced by SCAgent. 
Thus, the numerator is the number of correctly inferred testing traces, and the denominator is the total number of testing traces. 
We report Stage~2 proposal statistics separately, such as accepted channels and rejection reasons, because these quantities measure proposal quality rather than end-to-end inference accuracy.

\begin{table}[!t]
\centering
\caption{Comparison with prior iOS OS-level side-channel studies.}
\label{tab:prior_comparison}
\resizebox{\linewidth}{!}{
\begin{tabular}{lccc}
\toprule
Method & Target-event discovery & Channel discovery & Leakage analysis \\
\midrule
Zhang et al.~\cite{ios} & Manual & Manual & Traditional ML \\
Wang et al.~\cite{wang2023danger} & Manual & Keyword-guided semi-automatic & Trained time-series model \\
\ourmethod{} (Ours) & Agent-assisted automatic & LLM proposer--verifier & ROCKET + TabPFN \\
\bottomrule
\end{tabular}
}
\end{table}

\begin{table}[!t]
\centering
\caption{\small End-to-end attack accuracy on foreground app identification and website fingerprinting.
\SCAgentall{} uses all discovered channels, while \SCAgentnew{} excludes channels identified by prior work~\cite{ios, wang2023danger}.}
\resizebox{0.47\textwidth}{!}{
\begin{tabular}{ccc}
\toprule
Method & Foreground App & Website Fingerprinting \\
\midrule
Zhang et al.~\cite{ios} & 89.2\% & 68.5\% \\
Wang et al.~\cite{wang2023danger} & 94.1\% & 93.7\% \\
\SCAgentall{} (Ours) & \textbf{98.1\%} & \textbf{96.7\%} \\
\SCAgentnew{} (Ours) & 95.0\% & 90.1\% \\
\bottomrule
\end{tabular}
}
\label{tab:main}
\end{table}

\subsection{Main Results}
\label{subsec:main}
In this section, we present the end-to-end results of attacks generated by \ourmethod{} on two classic benchmarks: \emph{foreground app identification} and \emph{website fingerprinting}.
We compare our results against two representative baselines that focus on OS-level side channels on iOS~\cite{ios, wang2023danger}.

Specifically, we consider the following baselines.
(i) Zhang et al.~\cite{ios} manually identify a set of exploitable side channels and construct attack classifiers using traditional machine learning methods.
Their approach first transforms raw side-channel traces into Symbolic Aggregate approXimation (SAX) strings to obtain a Bag-of-Patterns (BoP) representation, and then trains an SVM classifier on the resulting features.
(ii) Wang et al.~\cite{wang2023danger} adopt a semi-automated pipeline that first identifies a candidate channel set using keyword-based heuristics, and then selects an optimal subset via mutual information analysis.
For modeling, they train a deep time-series classifier, specifically an InceptionTime~\cite{InceptionTime} model, from scratch.

Although the two baselines were published in different years, they remain the closest and most relevant prior iOS OS-level side-channel attacks to our setting, to the best of our knowledge. 
They study the same class of OS-exposed interfaces and evaluate closely related tasks, including foreground app identification and website fingerprinting. 
Therefore, they provide a meaningful comparison point for assessing whether \ourmethod{} improves over manually constructed or partially automated iOS side-channel pipelines.

\autoref{tab:prior_comparison} summarizes the main methodological differences. 
Unlike prior studies, \ourmethod{} automates both sensitive-event discovery and channel proposal within a unified pipeline, and further replaces per-task model training with a foundation-model-based few-shot analysis pipeline.

For \ourmethod{}, we evaluate two settings. 
In the \SCAgentall{} setting, the attack is constructed from channels discovered, verified, and selected by \ourmethod{}, including channels that may overlap with prior work. 
In the \SCAgentnew{} setting, we remove all channels reported by prior iOS side-channel studies before attack construction and use only the remaining discovered channels. 
Both settings run the complete discovery pipeline, including Stage~2 channel proposal and verification.

The \SCAgentnew{} setting models a stricter defensive scenario in which applications that aggressively query known vulnerable channels are filtered during vetting. 
It therefore evaluates whether \ourmethod{} can discover exploitable channels beyond the prior knowledge base, while \SCAgentall{} evaluates the best attack constructed by the full pipeline. 
In our experiments, only 4 of the 11 channels reported by prior iOS side-channel studies are selected in \SCAgentall{}, indicating that \ourmethod{} does not simply reuse the prior channel set.

The results are summarized in \autoref{tab:main}.
Overall, \SCAgentall{} achieves the highest accuracy on both benchmarks, indicating that the end-to-end pipeline produced by \ourmethod{} is highly effective.
For foreground app, \SCAgentall{} outperforms the strongest baseline, Wang et al.~\cite{wang2023danger}, by 4\%.
For website fingerprinting, it improves over the strongest baseline, Wang et al.~\cite{wang2023danger}, by 3\%.
These gains suggest that jointly automating side-channel discovery and attack construction, enabled by large language models and foundation models, allows \ourmethod{} to identify more informative channel combinations than prior approaches that rely on manually curated or partially automated pipelines.

More importantly, we observe that even under a constrained setting where channels identified by prior work are explicitly excluded, \SCAgentnew{} remains highly effective.
Across both benchmarks, \SCAgentnew{} preserves attack accuracy above 90\%, demonstrating that its performance is not solely driven by reusing previously known vulnerable channels.
In the case of foreground app identification, \SCAgentnew{} even slightly outperforms Wang et al.~\cite{wang2023danger} by 0.9\%, despite operating under stricter channel constraints.
This result indicates that \ourmethod{} is able to uncover previously unexplored side channels that are competitive with, and in some cases complementary to, known channels.
These findings suggest that defensive strategies based solely on filtering or monitoring known side channels may be insufficient.
Even when such channels are removed from consideration, \ourmethod{} can adapt by discovering alternative channels and constructing effective attacks, highlighting the need for more systematic defenses.

\para{Validation on recent iOS.}
We further evaluate whether the discovered leakage remains observable on iOS 26.4.2. 
\ourmethod{} achieves 96.5\% accuracy on website fingerprinting and 82.0\% accuracy on foreground app identification. 
The website-fingerprinting result is close to the main benchmark result, suggesting that the leakage is not specific to older iOS versions. 
The foreground-app result is lower, which is expected because the recent-iOS validation uses only four training traces and one testing trace per app due to manual collection constraints. 
Nevertheless, achieving 82.0\% accuracy over 100 apps under this low-data setting still indicates that the leakage remains observable on recent iOS.

\begin{figure*}[!t]
\centerline{\includegraphics[width=.95\linewidth]{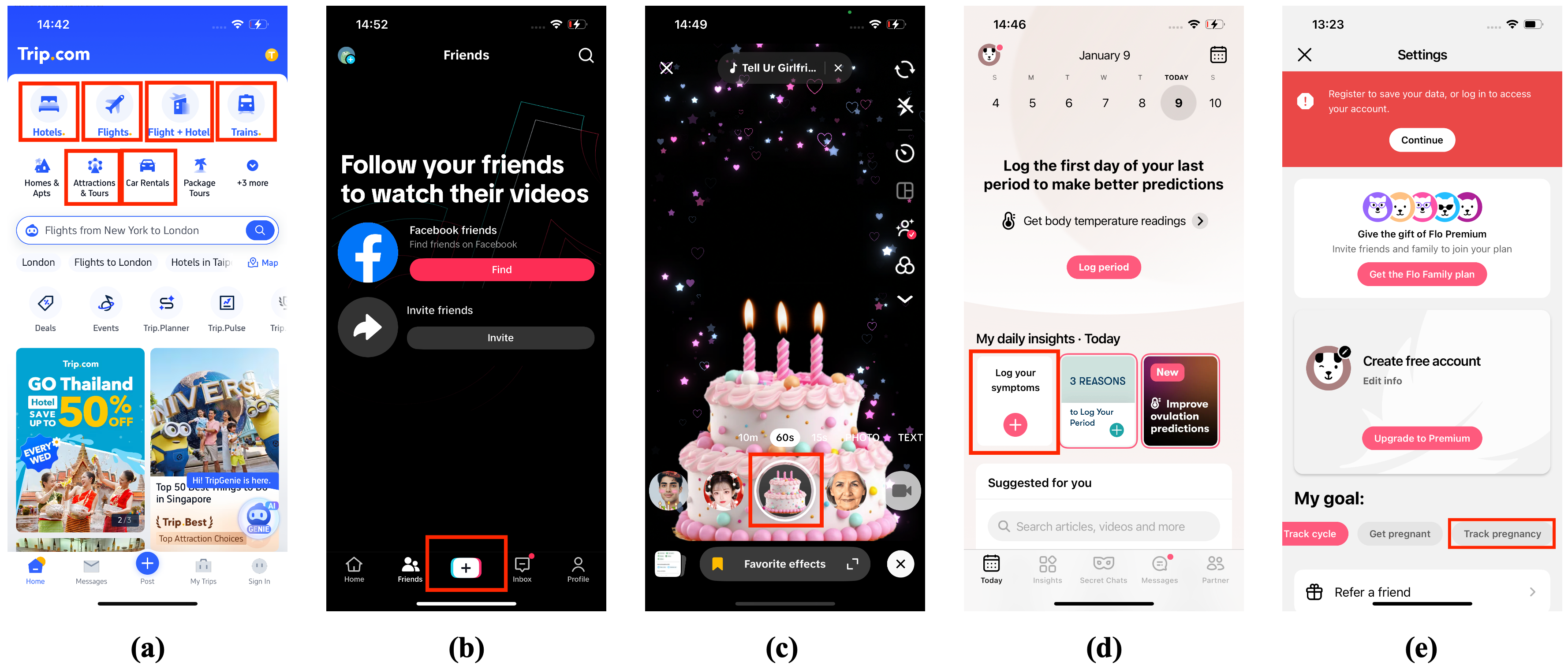}}
\caption{\small Identified sensitive in-app activities.
Sub-figure (a) shows the Trip app, where travel-related options are directly accessible from the home screen (highlighted by red boxes). By identifying clicks on these options, an attacker can infer users' travel-related activities.
Sub-figures (b) and (c) illustrate the TikTok app, where users first tap the ``Add'' button in (b) and then select a template in (c). The choice of templates may leak sensitive personal information, such as a user's birthday.
Sub-figures (d) and (e) depict the Flo app. In (d), users tap ``Log your symptoms'' to record health-related events. In the app settings shown in (e), users can enable ``Track pregnancy''. Such interactions and records may reveal critical personal states, such as pregnancy.}
\label{fig:in-app}
\end{figure*}

\subsection{Selected Channels}

In this section, we discuss representative \emph{new} side channels identified by \ourmethod{} that are most effective for the foreground app identification and website fingerprinting tasks.

Specifically, the selected channels include
(i) \texttt{fs\_free\_bytes},
(ii) \texttt{latency\_gpu},
(iii) \texttt{latency\_ane},
(iv) \texttt{latency\_filesystem},
(v) \texttt{latency\_cache}, and
(vi) \texttt{jitter\_us}.
Among these channels, \texttt{fs\_free\_bytes}, \texttt{latency\_gpu}, and \texttt{latency\_ane} are effective for both tasks.
The channel \texttt{jitter\_us} is primarily associated with foreground app identification, while \texttt{latency\_cache} and \texttt{latency\_filesystem} are specific to website fingerprinting.
All of these channels are automatically identified by \ourmethod{}.

The channel \texttt{fs\_free\_bytes} corresponds to the amount of free disk space reported by invoking \texttt{statfs("/private/var", \&stats)}.
This channel reflects transient changes in filesystem state caused by application- or browser-level activities, such as caching, logging, and temporary file creation.
Different foreground applications and websites induce distinct patterns of disk usage over time, making this signal informative for both tasks. Using only the \texttt{fs\_free\_bytes} channel, \ourmethod{} achieves 93.1\% accuracy on foreground app identification and 71.7\% on website fingerprinting.

\texttt{latency\_gpu} exploits contention on the GPU execution pipeline to infer ongoing GPU activity.
It measures GPU execution latency using \texttt{MTLCreateSystemDefaultDevice} by obtaining the default GPU device, submitting a lightweight workload, blocking the CPU thread until completion, and recording the resulting blocking time.
The observed latency serves as a proxy for GPU utilization and contention induced by different applications or web pages. Using only the \texttt{latency\_gpu} channel, \ourmethod{} achieves 82.4\% accuracy on foreground app and 81.6\% on website fingerprinting.

\texttt{latency\_ane} measures execution latency associated with the Apple Neural Engine (ANE) using \texttt{VNImageRequestHandler}.
By submitting a lightweight Vision request and measuring its execution time, this channel captures variations in ANE contention and utilization.
Different applications and web pages exhibit distinct ANE usage patterns, making this signal informative for both identification tasks.
Using only the \texttt{latency\_ane} channel, \ourmethod{} achieves 84.1\% accuracy on foreground app identification and 73.1\% on website fingerprinting.

\texttt{jitter\_us} exploits scheduler-induced timing variability caused by interactive foreground activity.
It measures fine-grained timing jitter using \texttt{mach\_absolute\_time}.
Under the preemptive scheduling model of the iOS kernel, background threads may be temporarily preempted when high-priority interactive tasks execute, introducing microsecond-level timing variability.
This scheduling interference makes \texttt{jitter\_us} informative for foreground app identification.
In contrast, website fingerprinting workloads are largely network- and I/O-driven and involve less consistent user interaction.
Consequently, the induced scheduling jitter is weaker and more variable, which limits the discriminative power of \texttt{jitter\_us} for distinguishing websites.
Using only the \texttt{jitter\_us} channel, \ourmethod{} achieves 81.7\% accuracy on foreground app identification.

\texttt{latency\_cache} measures cache access latency by repeatedly accessing a memory buffer with a size comparable to the L2 cache.
When the cache is lightly loaded, memory accesses exhibit low and stable latency.
In contrast, elevated latency indicates increased cache contention or congestion on the memory bus, reflecting interference from concurrent system activity.
Using only the \texttt{latency\_cache} channel, \ourmethod{} achieves 72.5\% accuracy on website fingerprinting.

\texttt{latency\_filesystem} measures filesystem synchronization latency using \texttt{fcntl(fd, F\_FULLFSYNC)}.
This operation forces pending file system writes to be flushed to stable storage, and the observed latency reflects the current I/O load and contention in the storage subsystem.
Variations in this latency are indicative of filesystem activity induced by different website loading behaviors.
Using only the \texttt{latency\_filesystem} channel, \ourmethod{} achieves 67.0\% accuracy on website fingerprinting.

We observe that among the channels, only \texttt{fs\_free\_bytes} is a \emph{passive} channel that reads an OS-exposed system state indicator (akin to reading resource counters from \texttt{/proc} on Linux).
In contrast, the remaining channels are \emph{active} side channels: the attacker deliberately issues specific system operations (e.g., GPU/ANE workloads, cache or filesystem probes) and infers the victim activity from the resulting latency or timing jitter.
Such active probing channels have received comparatively less attention in prior iOS side-channel studies, which predominantly focus on passively monitored OS-level statistics.
The complete list of newly identified channels is provided in \autoref{sec:Channels}.

Overall, these channels illustrate that \ourmethod{} can uncover a diverse set of side channels, including previously underexplored active channels arising from attacker-initiated probes.

\para{Clarifying channel novelty.}
We use the term ``new channel'' to refer to a concrete OS-level measurement primitive that has not been reported as exploitable in prior iOS OS-level side-channel studies under the same unprivileged-app threat model. 
This does not imply that the underlying physical phenomenon, such as contention or latency variation, is conceptually new. 
For example, cache- or latency-related leakage has been studied in other contexts, especially in microarchitectural side channels. 
However, the channels reported here are different in their attack surface: they are exposed through documented or observable OS-level interfaces, require no hardware-specific reverse engineering, and can be queried by an ordinary app within the standard iOS sandbox. 
Thus, the novelty lies in identifying previously unreported iOS-accessible measurement primitives and showing that their combination enables practical inference even when channels known from prior iOS work are excluded.

\begin{table}[!t]
\centering
\caption{End-to-end attack accuracy on in-app activities.}
\resizebox{0.47\textwidth}{!}{
\begin{tabular}{cccc}
\toprule
App & In-App Activity & \SCAgentall{} & \SCAgentnew{} \\
\midrule
Trip & \begin{tabular}[c]{@{}c@{}}hotels, flights, flight+hotel, trains,\\ attractions\&tours, car\_rentals\end{tabular} & 98.57\% & 92.86\% \\
\midrule
Flo & \begin{tabular}[c]{@{}c@{}}track\_pregnancy, track\_cycle,\\ log\_period, unprotected\_sex,\\ log\_ovulation\_test, link\_your\_partner\end{tabular} & 97.14\% & 94.29\% \\
\midrule
TikTok & \begin{tabular}[c]{@{}c@{}}aged\_filter, birthday\_cake,\\ CNY\_beauty, fire\_flames,\\ pink\_funny\_bando, happy\_birthday\end{tabular} & 98.33\% & 98.33\% \\
\bottomrule
\end{tabular}
}
\label{tab:in-app}
\end{table}

\subsection{Results on Sensitive In-App Activities}
We next extend our evaluation beyond classic benchmarks to sensitive in-app activities newly identified by \ourmethod{}. 
As described in \autoref{subsec:setup}, \ourmethod{} explores cross-platform apps using Mobile-Agent-E with Gemini~3~Pro, generating 1,433 candidate user actions and selecting 18 representative sensitive events for evaluation. 
These events cover different app categories and sensitivity types, including travel-related activities, health-related states, and profile-related information.

From the explored apps, we select three representative applications: Trip\footnote{\url{https://www.trip.com}}, Flo\footnote{\url{https://app.flo.health}}, and TikTok\footnote{\url{https://www.tiktok.com}}. 
Trip is a travel application that allows users to browse transportation options and itineraries; interactions with travel-related options, such as selecting trip types, may reveal travel plans and mobility patterns. 
Flo is a health-tracking application that enables users to log symptoms and personal health events; interactions within the app may expose reproductive-health information, including pregnancy-related records. 
TikTok is a widely used social media application for creating and sharing short-form videos; interactions such as selecting content-creation templates may reveal demographic or profile-related information, such as birthdays.

These applications cover diverse sensitivity types: private behavioral attributes (e.g., travel details), critical personal states (e.g., pregnancy), and profile-related or context-dependent personal attributes (e.g., birthdays). 
They also represent common app categories, including travel services, personal health tracking, and AI-powered content creation. 
Even when users may voluntarily share some information on a social platform, inferring the same information through a background monitoring app remains privacy-relevant because the inference occurs without user awareness, consent, or the visibility controls provided by the social application.

\autoref{fig:in-app} illustrates representative in-app activities for the three applications, and \autoref{tab:in-app} summarizes all monitored activities. 
For each application, we consider six distinct user activities, resulting in a six-class classification task.

The attack results are reported in \autoref{tab:in-app}. 
\SCAgentall{} consistently achieves test accuracy above 97\%, while \SCAgentnew{} maintains test accuracy above 92\% across all applications. 
These results show that the sensitive in-app activities automatically identified by \ourmethod{} correspond to practical and exploitable side-channel risks.

\begin{table}[!t]
\centering
\caption{Impact of the channels used in \ourmethod{}.}
\resizebox{0.47\textwidth}{!}{
\begin{tabular}{ccc}
\toprule
Channel & Foreground App & Website Fingerprinting \\
\midrule
Zhang et al.~\cite{ios} & 97.7\% & 91.9\% \\
Wang et al.~\cite{wang2023danger} & 97.8\% & 93.6\% \\
\SCAgentall{} (Ours) & \textbf{98.1\%} & \textbf{96.7\%} \\
\SCAgentnew{} (Ours) & 95.0\% & 90.1\% \\
\bottomrule
\end{tabular}
}
\label{tab:channel}
\end{table}

\begin{table}[!t]
\centering
\caption{Impact of the training method adopted in \ourmethod{}.}
\resizebox{0.47\textwidth}{!}{
\begin{tabular}{ccc}
\toprule
Method & Foreground App & Website Fingerprinting \\
\midrule
SVM & 71.0\% & 52.0\% \\
InceptionTime~\cite{InceptionTime} & 94.9\% & 96.2\% \\
TabPFN (Ours) & \textbf{98.1\%} & \textbf{96.7\%} \\
\bottomrule
\end{tabular}
}
\label{tab:method}
\end{table}

\subsection{Ablation Study}

In this section, we conduct an ablation study on \ourmethod{} to examine the contribution of key components. Specifically, we evaluate the effectiveness of the channels identified in Stage~2 and the impact of the training pipeline introduced in Stage~3.

\para{Impact of Channels.}
The results under different channel selections are reported in \autoref{tab:channel}. 
Specifically, we compare the channels discovered by \ourmethod{} with the channel sets used in prior work~\cite{ios, wang2023danger}. 
We further consider two variants: \SCAgentall{}, which exploits all discovered channels, and \SCAgentnew{}, which excludes channels previously studied in the literature. 
In this experiment, all other settings are kept identical to those of \ourmethod{}, and only the channel selection is varied.
We observe that \SCAgentall{} achieves the best performance among all channel selections, outperforming those used in prior work~\cite{ios, wang2023danger}. 
Notably, \SCAgentnew{} also attains performance comparable to that of prior work, despite relying solely on newly discovered channels.
These results demonstrate the effectiveness of the iterative hypothesis--verification framework in \ourmethod{}.

\para{Impact of Training Methodology.}
The results under different training methodologies are reported in \autoref{tab:method}. 
In this experiment, we fix the channel selection to those identified by \SCAgentall{} and vary only the training methodology to isolate its contribution. 
Specifically, we compare three approaches commonly considered in prior work: 
(i) a traditional machine learning method based on support vector machines (SVMs), 
(ii) InceptionTime~\cite{InceptionTime}, a deep neural network specifically designed for time-series classification, 
and (iii) the tabular foundation model TabPFN~\cite{TabPFN_Nature} adopted in \ourmethod{}.

We observe that, although TabPFN is not explicitly designed for time-series data, when combined with our ROCKET-based feature extraction~\cite{ROCKET}, it consistently outperforms time-series--specific methods. 
For example, on the foreground app identification task, TabPFN outperforms InceptionTime by 3.2\%, which constitutes a substantial improvement.

\begin{table}[!t]
\centering
\caption{Impact of the LLMs adopted in \ourmethod{}.}
\resizebox{0.43\textwidth}{!}{
\begin{tabular}{ccccccc}
\toprule
{Model} & {BER} & {SE} & {LD} & {CD} & {$\left| \mathcal{C}_{\text{db}} \right|
$} & {TP} \\
\midrule
Gemini 3 Pro & 7/80 & 3/80 & 30/80 & 1/80 & 39/80 & 80 \\
GPT‑4o & 5/60 & 9/60 & 10/60 & 12/60 & 34/60 & 60 \\
Grok 3 & 3/50 & 9/50 & 15/50 & 3/50 & 19/50 & 50 \\
\bottomrule
\end{tabular}
}
\label{tab:LLM}
\end{table}

\subsection{Impact of LLM Choice}
\label{sec:llm_choice}
In this section, we study how the choice of large language models (LLMs) in Stage~2 of \ourmethod{} affects channel discovery, focusing on the selection of the proposal model $M_p$ and the verification model $M_v$. 
By default, both $M_p$ and $M_v$ are instantiated using Gemini~3~Pro. 
We additionally evaluate two alternative models, GPT-4o and Grok~3, while keeping the prompt fixed (see \autoref{sec:Prompts}).

Recall that candidate channels proposed by $M_p$ are subsequently verified by $M_v$, and only those that pass verification are admitted into the discovered channel set $\mathcal{C}_{\text{db}}$. 
Proposals are generated iteratively in batches of size 10 by default, and the discovery process terminates automatically when the stopping criterion $\tau_{\text{stop}} = 0.2$ is met. We denote by $\mathit{TP}$ the total number of proposals generated prior to automatic termination.

We observe that rejected proposals can be broadly categorized into four failure types:
(i) \emph{Background Execution Restrictions (BER)}: although the prompt explicitly asks for channels that can be exploited when the app runs in the background (e.g., by masquerading as a music player), the LLM may hallucinate such capabilities and propose channels that are in fact unavailable under background execution. 
(ii) \emph{Sandbox Enforcement (SE)}: some proposals violate iOS’s mandatory sandboxing and resource isolation mechanisms, again reflecting hallucinated capabilities; we categorize these failures as SE.
(iii) \emph{Low Distinguishability (LD)}: as described earlier, we employ a lightweight per-channel usability assessment to evaluate whether a channel exhibits measurable event-related distinguishability. Channels that fail to meet the threshold $\tau_{\text{dist}}$ are classified as LD.
(iv) \emph{Channel Duplication (CD)}: since $M_p$ generates proposals in a batch-wise manner, it may repeatedly propose semantically identical or highly similar channels, which we label as CD.

The results are summarized in \autoref{tab:LLM}. 
We observe that Gemini~3~Pro terminates latest, generating a total of $\mathit{TP}=80$ proposals before auto-stopping, whereas GPT-4o and Grok~3 generate only 60 and 50 proposals, respectively. 
This suggests that Gemini~3~Pro is able to sustain a higher rate of valid proposals across iterations, indicating overall higher proposal quality compared to GPT-4o and Grok~3.
Consistent with this observation, Gemini~3~Pro also yields the largest set of exploitable channels, with $\lvert \mathcal{C}_{\text{db}} \rvert = 39$, while GPT-4o and Grok~3 produce 34 and 19 exploitable channels, respectively. 
Notably, although Gemini~3~Pro generates more proposals in total, its acceptance rate remains competitive, implying that its later-stage proposals continue to pass verification rather than being dominated by trivial or invalid suggestions.

Analyzing the failure modes in more detail, Gemini~3~Pro exhibits the highest number of low-distinguishability (LD) failures, which is expected given its larger proposal volume and broader exploration of the channel space. 
In contrast, Grok~3 produces the largest number of channel duplication (CD) cases, suggesting a tendency to repeatedly propose semantically similar channels under batch-wise generation.
All evaluated models produce a non-negligible number of failures due to background execution restrictions (BER) and sandbox enforcement (SE), reflecting inherent hallucination issues when reasoning about OS-level constraints.

Overall, these results highlight that hallucination remains a fundamental challenge for LLM-driven channel discovery, even for the state-of-the-art models. 
They further demonstrate the necessity of the verification model $M_v$ and the iterative  hypothesis--verification pipeline in \ourmethod{}, which effectively filters invalid or non-exploitable channels and prevents hallucinated capabilities from contaminating the final channel set.

\begin{figure}[!t]
\centerline{\includegraphics[width=.9\linewidth]{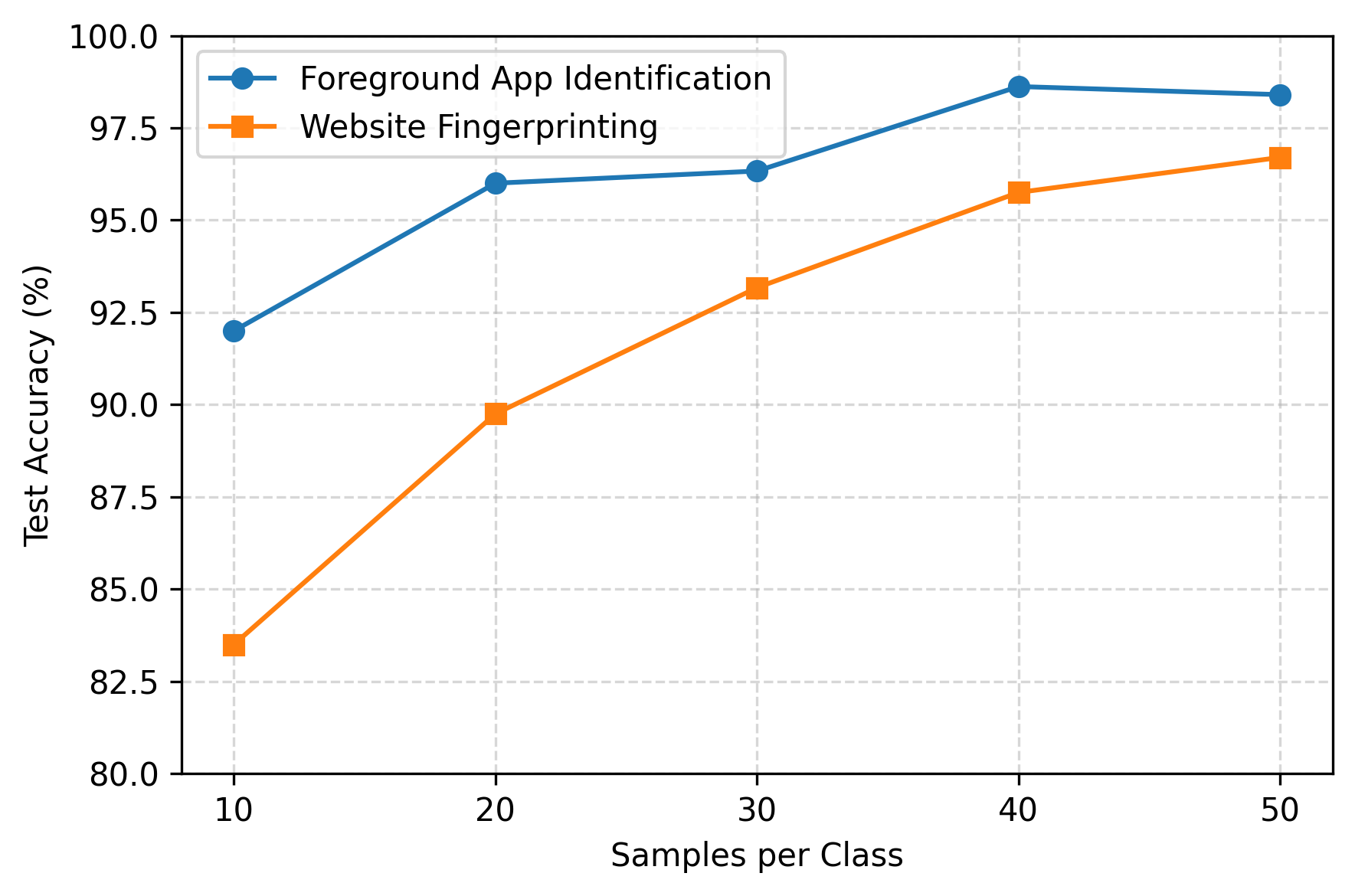}}
\caption{Impact of the dataset size collected in \ourmethod{}.}
\label{fig:datasize}
\end{figure}

\subsection{Impact of Dataset Size}
In this section, we evaluate the impact of dataset size on the effectiveness of \ourmethod{}. 
By default, each target event is triggered 50 times, yielding $N=50$ samples per event. 
The collected samples are split into training and testing sets using a fixed ratio of $\alpha = 0.8$. 
To study data efficiency, we progressively reduce the dataset size and evaluate $N \in \{10, 20, 30, 40, 50\}$, while keeping the split ratio unchanged. 
We consider two representative benchmarks for evaluation: foreground app identification and website fingerprinting.

The results are summarized in \autoref{fig:datasize}. 
As expected, reducing the number of samples generally leads to lower test accuracy, reflecting the increased difficulty of learning reliable decision boundaries under limited data. 
Nevertheless, \ourmethod{} remains highly effective even in low-data regimes. 
In particular, when the number of samples per event is reduced to $N=20$, the attacker's test accuracy still remains close to or above 90\% across both benchmarks.

These results indicate that \ourmethod{} is not overly sensitive to dataset size and can achieve strong performance with relatively few samples per event. 
This robustness can be attributed to two factors. 
First, the selected side channels exhibit strong event-related signals, which reduces the sample complexity required for learning. 
Second, the use of foundation model-based analysis enables effective generalization even when training data are scarce.

\begin{table}[!t]
\centering
\caption{Impact of device and threat model.}
\resizebox{0.4\textwidth}{!}{
\begin{tabular}{ccc}
\toprule
Device & \SCAgentall{} & \SCAgentnew{} \\
\midrule
iPhone 14 (White-box) & 96.7\% & 90.1\% \\
iPhone 14 (Black-box) & 95.3\% & 86.0\% \\
iPhone 13 (Black-box) & 95.0\% & 89.6\% \\
\bottomrule
\end{tabular}
}
\label{tab:device}
\end{table}

\subsection{Cross-Device Generalization under Black-Box Settings}

In the previous sections, we reported results under a white-box setting on iPhone~14, where the attacker is assumed to know the victim's device model and therefore collects training data exclusively on the same device type. 
In this section, we relax this assumption and evaluate the cross-device generalization of \ourmethod{} under a more realistic black-box setting. 
Specifically, we consider iPhone~13 and iPhone~14, where the attacker does not know in advance which device will be used by the victim and thus trains a single attack model using data collected from both devices. 
We focus on the website fingerprinting task in this experiment.

The results are shown in \autoref{tab:device}. 
Compared to the white-box setting, moving to the black-box setting leads to a modest reduction in attack accuracy: approximately 1\% when all exploitable channels are available, and close to 2\% on average when only newly discovered channels are used. 
Despite this degradation, \ourmethod{} remains highly effective, demonstrating that the learned attack models generalize well across device models.

We further observe that the attack performance is broadly comparable across iPhone~13 and iPhone~14. 
Under the \SCAgentall{} setting, slightly higher accuracy is observed when the attack is evaluated on iPhone~14. 
In contrast, under the \SCAgentnew{} setting, the attack achieves higher accuracy when evaluated on iPhone~13.
This asymmetry likely reflects device-specific differences in hardware characteristics and OS-level behaviors that influence the strength of individual side channels, rather than fundamental limitations of the attack.
Overall, these results indicate that \ourmethod{} does not rely on device-specific overfitting and can successfully transfer across closely related hardware platforms. 
This transferability significantly strengthens the practicality of the attack.

%% file: 5_discussion.tex
\section{Discussion}
\label{sec:discussion}

\begin{figure}[!t]
\centerline{\includegraphics[width=.9\linewidth]{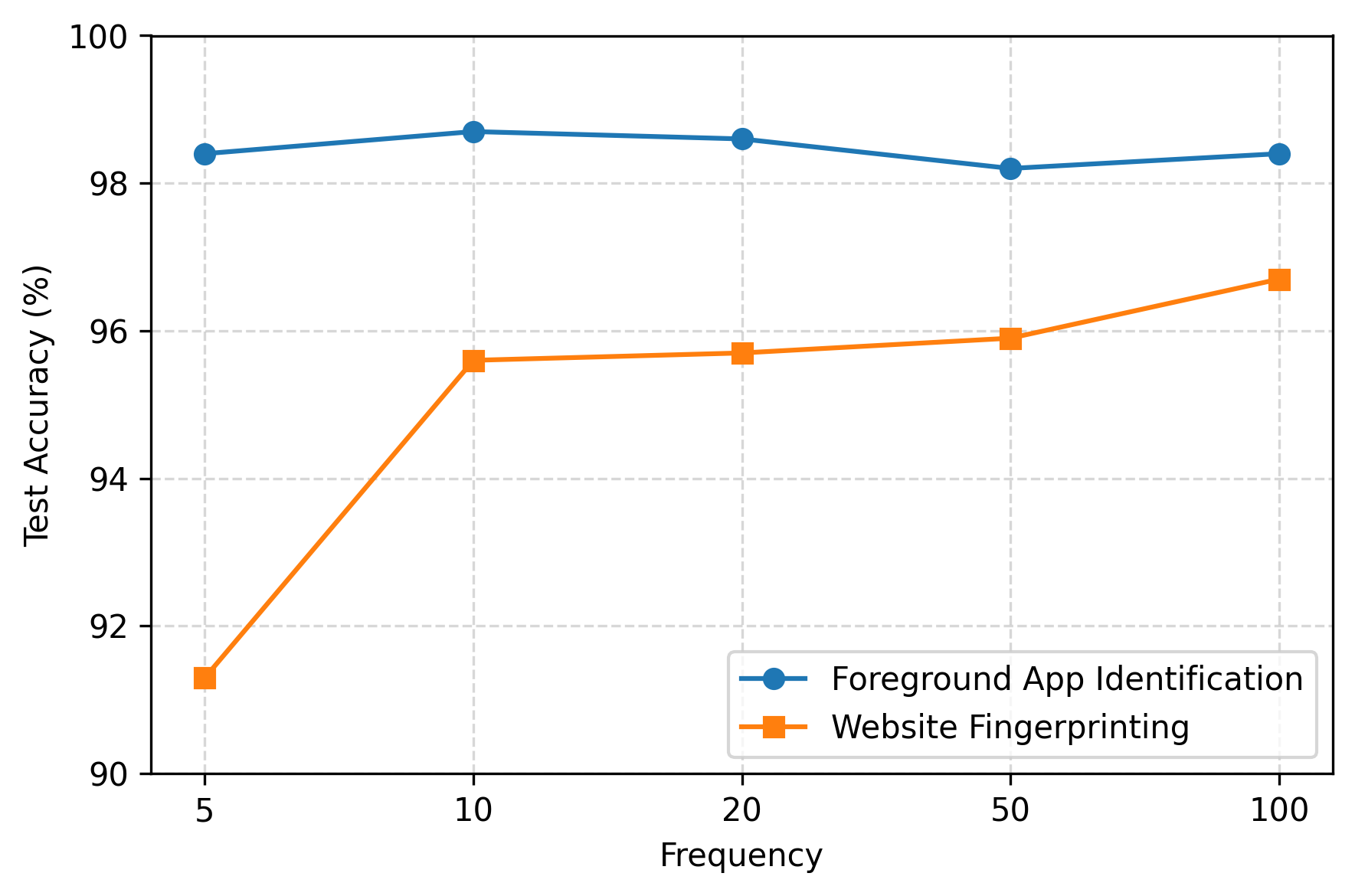}}
\caption{Performance of \ourmethod{} against frequency reduction.}
\label{fig:frequency}
\end{figure}

\begin{figure}[!t]
\centerline{\includegraphics[width=.9\linewidth]{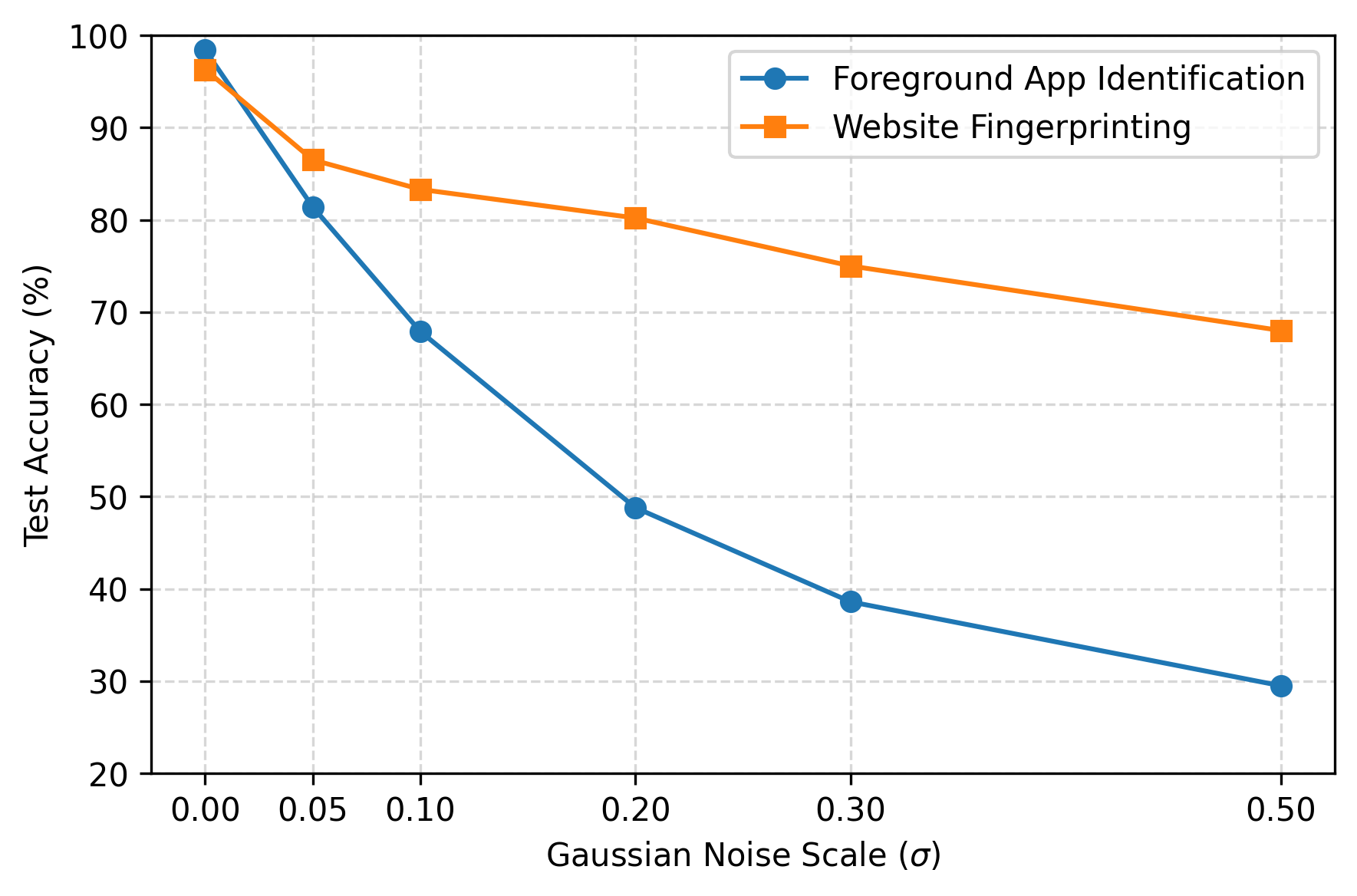}}
\caption{Performance of \ourmethod{} against gaussian noise.}
\label{fig:noise}
\end{figure}

\subsection{Potential Countermeasures}

In this section, we investigate potential countermeasures against \ourmethod{}. 
Recall that in the previous section, we already evaluated a defensive setting in which previously identified channels are monitored and aggressively queried APIs are detected during vetting. 
Here, we consider two additional classes of defenses:
(i) \emph{frequency reduction}, where the system proactively limits the refresh rate of API outputs, such that even high-frequency queries do not yield additional information beyond a capped temporal resolution; and 
(ii) \emph{Gaussian noise addition}, where APIs inject small, unnoticeable noise into returned values.
In the latter case, we assume that Gaussian noise with standard deviation $\sigma$ is added, which is feasible in practice since benign applications typically do not require high-precision readings.
We evaluate these defenses on two benchmarks: foreground app and website fingerprinting.

\para{Frequency reduction.}
The results are summarized in \autoref{fig:frequency}. 
We observe that reducing the refresh rate has little impact on foreground app identification. 
Even when the frequency is reduced to 5, which can be considered a normal and benign querying rate, the attack still achieves a test accuracy above 98\%. 
For website fingerprinting, reducing the frequency to 5 leads to a more noticeable degradation, with attack accuracy dropping by nearly 6\%. 
While this indicates that frequency reduction can partially mitigate leakage for this task, the attack accuracy remains above 90\%, suggesting that substantial information leakage still persists.

\para{Noise addition.}
The results are summarized in \autoref{fig:noise}. 
As expected, increasing the noise magnitude generally reduces the attacker’s test accuracy. 
When the noise standard deviation is set to $\sigma = 0.5$, the test accuracy for website fingerprinting drops to nearly 30\%, indicating a significant reduction in attack effectiveness, although some leakage remains given that random guessing yields an accuracy of approximately 1\%. 
Importantly, this defense comes at a non-trivial utility cost, as adding noise directly degrades the fidelity of API outputs and may adversely affect legitimate applications that rely on accurate measurements.

\para{Defensive implications.}
Our results suggest that defending only known interfaces is insufficient, because \ourmethod{} can discover alternative channels that remain exploitable after previously reported channels are excluded. 
A more robust defense should combine interface-level and subsystem-level mitigation. 
Interface-level defenses include reducing the precision or refresh rate of API outputs and monitoring abnormal query patterns during app vetting. 
Subsystem-level defenses aim to reduce shared-resource interference at the source, for example by isolating resource accounting, batching updates, or exposing coarser-grained system states to third-party apps. 
Passive defenses, such as API noise addition, can reduce leakage but may degrade benign utility, while active defenses, such as detecting repeated probing across multiple APIs, can target attack behavior more directly. 
Overall, \ourmethod{} highlights the need for systematic leakage auditing rather than one-off patching of known channels.

\subsection{Cost and Overhead Analysis}

In this section, we analyze the practical costs and overheads of \ourmethod{}, including LLM API usage costs, execution time, and power consumption. 
These factors are critical for assessing the feasibility of the attack in real-world settings.

\para{LLM API costs.}
Our channel discovery pipeline relies on commercial LLM APIs, incurring a small monetary cost.
Specifically, the discovery of sensitive actions costs \$1.98 per app.
During channel exploration, the candidate generator incurs a cost of \$0.18 per batch, while the leakage verifier incurs an additional \$0.04 per batch.
Overall, the total LLM API cost is modest, especially considering that channel discovery is a one-time or infrequent process per app.
This suggests that the economic barrier for mounting the attack is low.

\para{Execution time.}
We further report the execution time of each stage in \ourmethod{}.
Discovering sensitive actions takes 3445.65~seconds per app.
The candidate generation stage requires 117.33~seconds per batch, while leakage verification takes 34.94~seconds per channel (including distinguishability assessment).
Finally, the side-channel analysis stage, including model training, takes 129.60~seconds.
These results indicate that while the initial discovery phase is time-consuming, the remaining stages incur moderate overhead and are practical for offline or semi-online attack scenarios.

\para{Power consumption.}
We evaluate the power consumption of our monitoring app while collecting side-channel traces on an iPhone~14 running iOS~16.3.
During the experiment, the monitoring app runs in the foreground and invokes APIs at a frequency of 100.
The measured power consumption of our app is approximately 0.31~W, which is substantially lower than that of popular applications such as Spotify, YouTube, and Amazon, whose power consumption typically exceeds 1.0~W under normal usage.
This result suggests that \ourmethod{} is unlikely to be detected through simple power-consumption-based vetting or monitoring, further underscoring its stealthiness.

\para{Human intervention.}
Human involvement in \ourmethod{} is limited to experimental setup, including app installation, account preparation when required, and mapping Android-discovered semantic events to their corresponding iOS workflows for validation. 
The generation of candidate actions, semantic filtering of sensitive events, proposal of candidate channels, verifier feedback, channel selection, and leakage analysis are executed automatically by the pipeline. 
In particular, the verifier rejects infeasible proposals according to documented interface behavior, sandbox constraints, background-execution constraints, and lightweight distinguishability checks, rather than through manual inspection of each proposed channel.

\subsection{Limitations and Future Directions}

\para{Cross-platform semantic discovery.}
\ourmethod{} uses Android-based mobile agents to discover candidate sensitive events and then validates the corresponding workflows on iOS before any side-channel measurement. 
This design leverages the shared user-facing semantics of popular cross-platform apps while keeping all channel discovery, trace collection, and leakage evaluation on iOS. 
Future work can further strengthen this stage by developing mature iOS-native exploration agents, enabling broader coverage of iOS-specific app behaviors.

\para{Device and OS versions.}
Our main large-scale measurements use jailbreakable iOS versions to support controlled trace collection; jailbreak is used only for measurement instrumentation and is not part of the attack. 
To further validate practical relevance, we also evaluate \ourmethod{} on iOS 26.4.2, where it achieves 96.5\% accuracy on website fingerprinting and 82.0\% accuracy on foreground app identification. 
Future longitudinal studies across more devices, iOS versions, and app versions would provide a more complete picture of how side-channel leakage evolves over time.

\para{Prompt structure.}
\ourmethod{} uses structured prompts to encode the threat model, feasibility constraints, and verification goals. 
These prompts specify the analysis constraints but do not enumerate the final discovered channels or event--channel pairs, which are generated and filtered by the proposer--verifier loop. 
Future work could further study prompt robustness across different LLM backends, templates, and verifier designs.

\para{Scope of applicability.}
This paper instantiates \ourmethod{} for OS-level side channels observable by unprivileged applications. 
The same workflow may also inform analysis of other leakage classes, but applying it beyond OS-exposed interfaces would require measurement primitives and feasibility checks tailored to the corresponding attack surface.

%% file: 6_conclusion.tex
\section{Conclusion}
\label{sec:conclusion}

In this paper, we presented \ourmethod{}, an automated framework for side-channel risk analysis that integrates agent-driven sensitive-event discovery, LLM-assisted channel discovery with explicit verification, and foundation-model-based leakage analysis. 
Instantiated on iOS, \ourmethod{} uncovers previously unreported OS-level side channels and shows that they can leak meaningful information about both standard benchmarks and newly identified sensitive in-app activities. 
More broadly, our results suggest that side-channel security should be treated as a systematic and ongoing auditing process, rather than as a collection of isolated manual case studies.

\section*{Ethical Considerations}

This study evaluates OS-level side-channel risks through controlled experiments on researcher-owned devices. 
Our experiments do not involve human subjects, user studies, or data collected from real users; all measurements are conducted using test accounts and researcher-generated interactions. 
The evaluated app activities and website visits are triggered only by the authors in a controlled laboratory setting, and the collected traces are used solely for measuring side-channel leakage under the threat model described in this paper.

We took several steps to minimize potential harm. 
First, we did not deploy the attack against any third-party users or devices. 
Second, we did not collect, store, or infer any personal data from real users. 
Third, our artifact is intended to support reproducibility and defensive analysis, and we avoid releasing materials that would enable unsupervised deployment against real users. 
Any examples involving sensitive activities are used only to demonstrate the type of information that could be exposed by OS-level leakage, not to profile or monitor actual individuals.

As part of the evaluation, we also examined whether an application capable of performing the monitored measurements would be flagged during app review. 
This test was conducted with safeguards: the application did not collect, transmit, or store user data. 
Although it passed the App Store review process, we did not proceed with the separate developer action required to release or list the application on the App Store. 
As a result, the application was never publicly available and received no downloads, so no users were exposed to the monitoring functionality.

Newly identified channels have been responsibly disclosed to the vendor prior to publication. 
We remain in communication with the vendor and present the results to raise awareness of systemic side-channel risks and to support the development of stronger platform-level mitigations.

%% file: main.bbl

%% file: 99_appendix.tex
\clearpage
\appendix

\subsection{Open Science}
The implementation of \ourmethod{} is available at: \url{https://anonymous.4open.science/r/SCAgent-D878/README.md}. The artifact includes the SCAgent implementation, prompt templates, channel measurement code, and metadata for the evaluated case studies. 
For each reported channel, we provide its measurement interface, required execution context, and verifier outcome. 
We do not release raw traces that may contain device- or app-specific behavioral information, but provide scripts and configuration files to reproduce the measurements on controlled devices.

\subsection{Generative AI Usage}
ChatGPT was used for minor grammar correction, language polishing, and writing assistance. All technical ideas, experimental designs, analyses, and conclusions were developed and verified by the authors.

\begin{algorithm}[!ht]
\caption{Automated Discovery of Side Channels}
\label{alg:stage2}
\begin{algorithmic}[1]
\Input Target event $E$, documentation $\mathcal{D}$, prior channels $\mathcal{C}_{\text{prior}}$, proposer $M_p$, verifier $M_v$, batch size $B$, thresholds $\tau_{\text{dist}}$, $\tau_{\text{stop}}$
\Output Candidate channel database $\mathcal{C}_{\text{db}}$

\State $\mathcal{C}_{\text{db}}^{(0)} \gets \mathcal{C}_{\text{prior}}$
\Comment{initialize database}
\label{alg:init-db}

\State $\mathcal{F}^{(0)}_{M_v} \gets \emptyset$
\Comment{initial verifier feedback}
\label{alg:init-feedback}

\State $t \gets 0$
\label{alg:init-iter}

\While{true}
    \State $\mathcal{C}_{\text{prop}}^{(t)} \gets 
    \mathcal{P}_{M_p}(E,\mathcal{D},\mathcal{C}_{\text{db}}^{(t)},\mathcal{F}^{(t)}_{M_v})$
    \Comment{batch proposal}
    \label{alg:propose}

    \State $\mathcal{C}_{\text{valid}}^{(t)} \gets \emptyset$
    \label{alg:init-valid}

    \State $\mathcal{F}^{(t+1)}_{M_v} \gets \emptyset$
    \label{alg:init-next-feedback}

    \For{$i = 1$ \textbf{to} $B$}
        \State $c \gets c^{(t)}_i$
        \label{alg:select-channel}

        \If{\textbf{not} $\textsc{SemanticCheck}(c,\mathcal{D},M_v)$}
            \State $\mathcal{F}^{(t+1)}_{M_v} \gets \mathcal{F}^{(t+1)}_{M_v} \cup \{c\}$
            \label{alg:reject-semantic}
            \State \textbf{continue}
        \EndIf

        \If{\textbf{not} $\textsc{ThreatCheck}(c,M_v)$}
            \State $\mathcal{F}^{(t+1)}_{M_v} \gets \mathcal{F}^{(t+1)}_{M_v} \cup \{c\}$
            \label{alg:reject-threat}
            \State \textbf{continue}
        \EndIf

        \If{$\mathrm{Acc}(c,E) \ge \tau_{\text{dist}}\cdot \mathrm{Acc}_{\text{rand}}(E)$}
            \State $\mathcal{C}_{\text{valid}}^{(t)} \gets \mathcal{C}_{\text{valid}}^{(t)} \cup \{c\}$
            \label{alg:accept-channel}
        \Else
            \State $\mathcal{F}^{(t+1)}_{M_v} \gets \mathcal{F}^{(t+1)}_{M_v} \cup \{c\}$
            \label{alg:reject-usability}
        \EndIf
    \EndFor

    \State $\mathcal{C}_{\text{db}}^{(t+1)} \gets \mathcal{C}_{\text{db}}^{(t)} \cup \mathcal{C}_{\text{valid}}^{(t)}$
    \Comment{database update}
    \label{alg:update-db}

    \State $\rho^{(t)} \gets |\mathcal{C}_{\text{valid}}^{(t)}| / B$
    \Comment{acceptance rate}
    \label{alg:accept-rate}

    \If{$\rho^{(t)} < \tau_{\text{stop}}$}
        \State \textbf{break}
        \Comment{stop condition}
        \label{alg:stop}
    \EndIf

    \State $t \gets t + 1$
    \label{alg:next-iter}
\EndWhile

\State \Return $\mathcal{C}_{\text{db}}^{(t)}$
\label{alg:return}
\end{algorithmic}
\end{algorithm}

\subsection{Algorithms}
\label{subsec:alg}

\autoref{alg:stage2} summarizes the automated discovery of side channels as an iterative, batch-based propose--verify loop. The process is initialized with side channels identified in prior work (line~\ref{alg:init-db}), ensuring that existing knowledge is treated as a starting point rather than a fixed boundary. At each iteration, the proposer $M_p$ generates a batch of candidate channels conditioned on the current database and verifier feedback (line~\ref{alg:propose}), enabling systematic exploration while avoiding redundant hypotheses.

Each proposed channel is evaluated by the verifier $M_v$ through a sequence of semantic, threat-model, and usability checks (lines~\ref{alg:reject-semantic}--\ref{alg:accept-channel}), filtering out hallucinated, infeasible, or uninformative channels before they enter the database. Verification outcomes are explicitly accumulated as structured feedback and fed back to the proposer in subsequent iterations (line~\ref{alg:init-next-feedback}), progressively improving proposal quality. The discovery process terminates when the acceptance rate of a proposal batch falls below a predefined threshold (line~\ref{alg:stop}), indicating diminishing returns from further exploration. This design allows \ourmethod{} to balance coverage and precision while maintaining scalability.

\autoref{alg:stage3} summarizes the automated side-channel analysis stage of \ourmethod{}. Given a target event and a set of verified candidate channels, the algorithm first selects a compact subset of channels tailored to the event (line~\ref{alg:stage3-select}), reducing measurement overhead and improving stealth. Side-channel traces collected from the selected channels are then transformed into a tabular representation via time-series feature extraction and selection (lines~\ref{alg:stage3-rocket}--\ref{alg:stage3-tabular}), enabling compatibility with tabular foundation models.

Rather than training a bespoke classifier, the algorithm applies TabPFN in an in-context learning manner using a labeled context set (lines~\ref{alg:stage3-split}--\ref{alg:stage3-fit}), allowing efficient evaluation under limited data. The resulting classifier and its predictive accuracy quantify the distinguishability of the target event from the observed side-channel signals. Together, this design enables scalable and automated side-channel analysis while avoiding the cost and fragility of per-channel model training.

\begin{algorithm}[!t]
\caption{Automated Side-Channel Analysis}
\label{alg:stage3}
\begin{algorithmic}[1]
\Input Candidate channel database $\mathcal{C}_{\text{db}}$, target event $E$, channel selector $M_s$, TabPFN model $\mathcal{T}$, feature extractor $\phi(\cdot)$, feature selector $\psi(\cdot)$, train/test split ratio $\alpha$
\Output Final classifier $\hat{f}$ and test accuracy

\State $\mathcal{C}_{\text{sel}} \gets \mathcal{S}_{M_s}(E,\mathcal{C}_{\text{db}})$
\Comment{select channels}
\label{alg:stage3-select}

\State $\mathcal{D} \gets \textsc{CollectTraces}(E,\mathcal{C}_{\text{sel}})$
\Comment{labeled MTS dataset}
\label{alg:stage3-collect}

\State $\Phi \gets \{\phi(X) \mid (X,Y)\in \mathcal{D}\}$
\Comment{ROCKET features}
\label{alg:stage3-rocket}

\State $\tilde{\Phi} \gets \{\psi(\phi(X)) \mid (X,Y)\in \mathcal{D}\}$
\Comment{reduced features}
\label{alg:stage3-reduce}

\State $\tilde{\mathcal{D}} \gets \{(\tilde{\phi}(X),Y)\}_{(X,Y)\in \mathcal{D}}$
\Comment{$\tilde{\phi}=\psi\circ\phi$}
\label{alg:stage3-tabular}

\State $(\tilde{\mathcal{D}}^{\text{train}},\tilde{\mathcal{D}}^{\text{test}}) \gets \textsc{Split}(\tilde{\mathcal{D}},\alpha)$
\Comment{context vs. test}
\label{alg:stage3-split}

\State $\hat{f} \gets \textsc{TabPFNClassifier}(\mathcal{T},\tilde{\mathcal{D}}^{\text{train}})$
\Comment{no fine-tuning}
\label{alg:stage3-fit}

\State $\widehat{Y}^{\text{test}} \gets \hat{f}(\tilde{\mathcal{D}}^{\text{test}})$
\Comment{ICL predictions}
\label{alg:stage3-predict}

\State $\mathrm{Acc} \gets \textsc{Accuracy}(\widehat{Y}^{\text{test}},Y^{\text{test}})$
\Comment{distinguishability metric}
\label{alg:stage3-acc}

\State \Return $(\hat{f},\mathrm{Acc})$
\label{alg:stage3-return}
\end{algorithmic}
\end{algorithm}

\clearpage

\onecolumn

\subsection{Prompts}
\label{sec:Prompts}

In this section, we present the prompts used in our experiments.

\begin{prettybox}{Prompt for Automated Sensitive Event Discovery:}
    \small

Role: Autonomous Side-Channel Vulnerability Mapper.
Target: Current App (trip.com).
Mission: Explore the app to identify UI elements that act as 'Side-Channel Sources' and explicitly state WHAT information they leak.

Constraint: Do NOT analyze CPU interrupts or keystroke timing. Focus strictly on Memory, Network, and GPU side-channels.
 Navigation Strategy (CRITICAL - BFS ENFORCEMENT):
 
  1. PRIORITIZE COVERAGE: Do not focus on a single vertical (e.g., Hotels) for more than 4 steps. You must switch categories frequently.
  
  2. BREADTH-FIRST SEARCH: If you see multiple top-level options (e.g., 'Flights', 'Hotels', 'Trains', 'Attractions'), you must explore a different one than the previous run.
  
  3. DEPTH LIMIT: Do not go deeper than the 'Details Page'. Once you identify a leak in a sub-category, explicitly perform a 'Back' action or tap 'Home' to explore a new category.
  
  4. ANTI-RABBIT HOLE: Do not fill out long forms. If you reach a form, log the risk and leave immediately to find a new button.
  
Use this 'UI-to-Leak' Logic Table for your exploration:

1. IF you see [Lists with Images / News / Search Results] $\to$ THEN conclude:
   - Vulnerability: Loading specific items creates unique Network Packet Size sequences.
   - Specific Leak: 'Exactly which item/article the user is interested in' (Traffic Analysis).

2. IF you see [Major Transitions (e.g., opening 'Payment', 'Camera', or 'Settings')] $\to$ THEN conclude:
   - Vulnerability: Activity switching alters Shared Memory \& WindowManager state.
   - Specific Leak: 'Current App Lifecycle State (e.g., User is about to pay vs. just browsing)' (UI State Inference).

3. IF you see [Infinite Scroll / Maps / Video / 3D Models] $\to$ THEN conclude:
   - Vulnerability: High-load rendering creates measurable GPU usage spikes.
   - Specific Leak: 'Specific complex content being viewed' (GPU Fingerprinting).

4. IF you see [Search Bars with Auto-complete] $\to$ THEN conclude:
   - Vulnerability: Real-time suggestions trigger network bursts per character.
   - Specific Leak: 'Approximate search query content' (via Traffic Analysis).

Execution \& Logging Requirement:
Explore autonomously. When you interact with a risky element, you MUST output a log in this exact format:
'RISK DETECTED: [UI Element Description] $\to$ LEAKS: [Specific User Secret] via [Side Channel Type]'

Example:

'RISK DETECTED: Hotel Result List $\to$ LEAKS: Specific Hotel Choice via Network Traffic Analysis'

'RISK DETECTED: Payment Activity $\to$ LEAKS: User Intent to Pay via UI State Inference'

You have to export the results within 40 steps.
Start exploring now.
\end{prettybox}

\begin{prettybox}{Prompt for the Channel Selector $M_s$: }
    \small 

Role
Act as an expert in Side-Channel Analysis and Hardware Fingerprinting.

Task
Based on the provided performance data (Accuracy) and technical descriptions, construct a robust fingerprinting classifier by selecting a diverse set of Side-Channel Types, with 1-2 specific vectors chosen per type to maximize reliability.

Selection Logic

1. Categorization (Diversity):

   First, classify all available vectors into distinct Hardware/Resource Types (specifically: Network I/O, Hardware Acceleration/GPU, Micro-architecture/Execution, and Memory/Storage).

2. Intra-Type Selection (Enhancement):

   For each identified Type, select the optimal configuration:
   
   - Primary Vector: Pick the single vector with the highest individual accuracy within this type.
   
   - Auxiliary Vector (Optional): Select a second vector within the same type ONLY IF it enhances the Primary Vector by measuring a different dimension (e.g., Static vs. Dynamic, Input vs. Output) or covers a specific blind spot (like AI usage).
   
   - Constraint: Do not exceed 2 vectors per type.

Goal
Propose a final list of channels that offers the best balance of High Accuracy and Low Correlation (Orthogonality). Explain the "Selection Logic" for each choice.
\end{prettybox}
 
\vspace{1em}

\begin{prettybox}{Prompt for the Proposer $M_p$:}
    \small 

Role Definition

You are an elite Security Researcher specializing in Browser-Based Side-Channel Attacks and Operating System Internals (specifically Apple XNU/iOS Sandbox). You have deep expertise in Website Fingerprinting (WF), WebKit architecture, Mach Kernel APIs, and hardware micro-architectural leaks.

 Context

I am providing you with the text/content of several academic papers regarding side-channel attacks.

 Your Mission

Your primary goal is TARGETED DISCOVERY for Website Fingerprinting. 
You must synthesize theoretical knowledge from the papers to identify specific side-channel vectors within the iOS Sandbox that can leak which website a user is currently visiting. 

Core Hypothesis: Different websites trigger unique resource consumption patterns (CPU bursts, GPU rendering loads, Neural Engine activation for image processing, Memory compression). You need to find ways to measure these patterns from a sandboxed app.

 Step-by-Step Execution Plan

 Step 1: Deconstruct \& Browser-Mapping (Theoretical Foundation)

Analyze the provided papers. Perform "Mechanism-to-Browser Mapping":

1.  Core Principle Extraction: What is the root cause of the leak? (e.g., "Shared Cache Contention", "Interrupt Latency", "DRAM Rowhammer").
2.  WebKit Contextualization: How does a modern browser (Safari/WebKit) trigger this resource?
     Example: If a paper discusses "GPU Interrupts," ask: "Does complex CSS/WebGL on a website cause measurable GPU interrupt jitter?"
     Example: If a paper discusses "DRAM bank conflict," ask: "Does the JavaScript JIT engine create detectable memory access patterns?"

 Step 2: EXTREME ATTACK SURFACE MAPPING (The "Hunt")

Conduct a BROAD yet FOCUSED search for side-channels that correlate with Page Loading \& Rendering.
CRITICAL REQUIREMENT: Use 'Google Search' to verify API existence and known browser behaviors.

1.  Systematic Subsystem Scan (Web-Centric Matrix):
    Focus on subsystems that fluctuate heavily during web browsing. You must search for vectors in EACH category:

     Category A: The Rendering Pipeline (Visual Fingerprints)
       GPU \& Metal: Search for "IOSurface creation overhead", "CAMetalLayer frame timing jitter", "detecting GPU context switch latency iOS". Can we infer the complexity of the DOM/CSS being rendered?
       Compositor (RenderServer): Search for "transaction commit latency", "CoreAnimation frame drops detection".

     Category B: Compute \& Intelligence (Logic Fingerprints)
       Neural Engine (ANE): Modern browsers use AI for "Live Text" (OCR) or image analysis. Search for "ANE locking side channel", "CoreML inference blocking main thread". Can we detect if a site has many images containing text?
       CPU \& Scheduler: Search for "QoS level switching latency", "detecting high-performance core activation".

     Category C: Memory \& Storage (Resource Fingerprints)
       Memory Pressure: Search for "jetsam event monitoring", "memory compression latency side channel". Large sites trigger memory compression; can we measure the compression daemon's activity?
       File System/Cache: Search for "WebKit disk cache side channel", "statfs blocking time".

     Category D: Media \& Peripherals (Content Fingerprints)
       Audio Stack: Search for "CoreAudio buffer underrun detection", "AudioSession interruption listener". Does the site auto-play video?

2.  Feasibility Check:
     For every vector, ask: "Is this visible to a background/foreground sandboxed app while Safari is active?"

\ Step 3: Comprehensive Inventory \& Selection

Output the Full Inventory: List ALL valid NEW side-channels, specifically explaining WHY this vector helps classify webs.

Selection Strategy:

 Generate 10 NOVEL distinct vectors which has never been mentioned in the papers that capture different dimensions of app activity.

 Priority: Favor vectors that can distinguish between architecturally similar apps (e.g., distinguishing TikTok from Instagram, not just "Game vs. Notes").
\end{prettybox}

\begin{prettybox}{Prompt for the Verifier $M_v$:}
    \small 

Role \& Objective

You are an expert iOS Security Researcher and Red Teamer specializing in side-channel analysis, OS internals (XNU/Darwin), and mobile privacy.

We are conducting a multi-turn feasibility study on potential Side-Channel Attack Vectors. I will provide vectors (APIs, behaviors, or resource constraints). Your job is to rigorously evaluate them on a non-jailbroken, modern iOS device.

 Core Analysis Logic (The "New Rules")

You must apply these two advanced logic filters to every evaluation:

 1. The "Inference via Latency" Rule

Do not stop at "Access Denied". If a global resource is blocked or returns generic errors, you must verify if Timing Side-Channels are possible.

- Question: Does the time it takes to fail (or return a result) change based on the system state?
- Key Indicator: Look for Shared Resource Contention (e.g., Lock Contention, Cache Eviction, Service Busy Status).

 2. The "Resource Bottleneck" Rule (Classification Standard)

Classify vectors by underlying physical constraint, not API names.

- Gold Standard for Differentiation:
  - Different Channel: Vector A triggers Disk I/O + IPC (e.g., 'UIFont.systemFont' loading a file) vs. Vector B triggers Memory Table Lookup (e.g., 'UIFont.familyNames' reading a cache). Even if they are in the same Framework, the bottleneck is different.
  - Same Channel: Vector A and Vector B run different code loops but both purely saturate the ALU/Int Unit. They are the SAME channel (Redundant).

 The "Negative Knowledge Base" (Baseline Failure Modes)

Use this checklist to identify known blockers.

1. Background Restrictions \& State-Based Blocking:

   - Scope: Hardware (Mic/Cam) or signals disconnected/silenced by the OS when suspended/backgrounded.
   
2. Sandbox \& Resource Isolation:

   - Scope: 'audit\_token' checks, Container isolation, specific XPC filtering.
   
3. Low Accuracy \& Physical Limitations:

   - Scope: Thermal throttling, signal-to-noise ratio too low.
   
4. Redundancy \& Resource Overlap (The "Duplicate" Check):

   - Scope: The vector offers no unique signature. It relies on the exact same hardware unit (e.g., ALU, L1 Cache) as a generic baseline or a previously discussed vector.

 Critical Constraints (Thinking "Outside the Box")

You must explicitly check for failure modes NOT listed above:

- Privacy Manifests: Does this API require a declared reason in 'Info.plist' (triggering Apple Review)?
- TCC \& Dynamic Permissions: Does it trigger a user prompt?
- Modern Mitigations: iOS specific patches.

 Response Protocol

For every vector, output the following structured analysis:

1. Feasibility Score: 0/10 (Impossible) to 10/10 (Confirmed Working).
2. The Blocker:
   - Cite Category 1-5, OR "OUTSIDE CONTEXT: [Reason]" (e.g., Entitlements, TCC).
3. Side-Channel Potential (Timing/Inference):
   - Analysis: Even if direct access fails, can we infer global state via latency/contention? (Yes/No + Explanation).
4. Mechanism \& Bottleneck (Signature):
   - Resource Type: Define the physical constraint (e.g., "Heavy I/O + IPC", "Pure ALU", "Syscall Context Switch", "Memory Bandwidth").
   - Differentiation: Does this logically differ from a standard CPU loop or previous vectors? (Refer to the Font I/O vs Memory example).
5. Technical Nuance:
   - Brief explanation of internal behavior.

System Initialized. Ready for Vector 1.
\end{prettybox}

\subsection{Summary of New Channels Discovered by \ourmethod{}}
\label{sec:Channels}
\renewcommand{\arraystretch}{1.3}

\begin{table}[!ht]
\centering
\begin{tabular}{|m{1.5cm}|m{5cm}|m{7cm}|}
\hline
\textbf{Number} & \textbf{Channel} & \textbf{API} \\ \hline

1 & vm\_compress & \multirow{11}{*}{host\_statistics64} \\ \cline{1-2}
2 & vm\_decompress & \\ \cline{1-2}
3 & vm\_comp\_pages & \\ \cline{1-2}
4 & vm\_uncomp\_pages & \\ \cline{1-2}
5 & vm\_ext\_pages & \\ \cline{1-2}
6 & vm\_int\_pages & \\ \cline{1-2}
7 & vm\_spec & \\ \cline{1-2}
8 & vm\_purges & \\ \cline{1-2}
9 & vm\_purgeable & \\ \cline{1-2}
10 & vm\_react & \\ \cline{1-2}
11 & vm\_pageouts & \\ \hline

12 & lo0\_ib & \multirow{3}{*}{getifaddrs} \\ \cline{1-2}
13 & lo0\_ob & \\ \cline{1-2}
14 & lo0\_ip & \\ \hline

15 & latency\_filesystem & \multirow{3}{=}{Virtual File System Syscalls (open / write / close / statfs / fread)} \\ \cline{1-2} 
16 & fs\_free\_bytes & \\ \cline{1-2}
17 & probeDyldWarmup & \\ \hline

18 & probe\_font\_family & \multirow{2}{*}{UIFont} \\ \cline{1-2}
19 & probe\_font & \\ \hline

20 & probe\_jit\_sim & UnsafeMutablePointer.allocate \\ \hline
21 & probe\_dns & gethostbyname \\ \hline
22 & jitter\_us & Native CPU Instructions (Swift Stdlib) \\ \hline
23 & probe\_img\_io & CGColorSpaceCreateDeviceRGB \\ \hline
24 & probe\_fs\_meta & FileManager.attributesOfItem \\ \hline
25 & probe\_socket\_latency & socket \\ \hline
26 & probePreferences & UserDefaults.standard \\ \hline
27 & latency\_cache & Native Array Access (Swift Stdlib) \\ \hline
28 & probeSpeechPrewarm & AVSpeechSynthesisVoice \\ \hline
29 & act\_main\_pre & DispatchQueue.main.sync \\ \hline
30 & latency\_gpu & MTLCommandQueue \\ \hline
31 & latency\_ane & VNImageRequestHandler \\ \hline
32 & latency\_aud\_blk & AVAudioSession \\ \hline
33 & phy\_acc\_z & CMMotionManager \\ \hline
34 & probeTCCD & Security.framework.SecItemCopyMatching \\ \hline
35 & probeNetworkStack & URLSession \\ \hline
36 & probeSpotlight & CSSearchableIndex \\ \hline
37 & probeQuickLook & QLThumbnailGenerator \\ \hline
38 & probeEventKit & EKEventStore \\ \hline
39 & probeSpeechPrewarm & AVSpeechSynthesisVoice \\ \hline

\end{tabular}
\label{tab:Channels}
\end{table}